\documentclass[11pt,a4paper]{article}
\usepackage{amsmath}
\usepackage{bm}
\usepackage{amssymb}
\usepackage{mathrsfs}
\usepackage{amsbsy}
\newcommand{\beq}{\begin{equation}}
\newcommand{\eeq}{\end{equation}}
\newcommand{\bea}{\begin{eqnarray}}
\newcommand{\eea}{\end{eqnarray}}
\newcommand{\bean}{\begin{eqnarray*}}
\newcommand{\eean}{\end{eqnarray*}}
\newcommand{\spinor}{{\mathcal S}}
\newcommand{\Spinor}{{{\pmb{\mathcal{S}}}}}
\newcommand{\bpm}{\begin{pmatrix}}
\newcommand{\epm}{\end{pmatrix}}

\newcommand{\da}{{\dot a}}
\newcommand{\db}{{\dot b}}
\newcommand{\dc}{{\dot c}}

\newcommand{\tr}{{T}}
\newcommand{\HatT}{{\widehat T}}
\newcommand{\HatP}{{\widehat P}}
\newcommand{\HatC}{{\widehat C}}
\newcommand{\bT}{{\bm T}}
\newcommand{\bP}{{\bm P}}
\newcommand{\bU}{{\bm U}}
\newcommand{\bV}{{\bm V}}
\newcommand{\bW}{{\bm W}}
\newcommand{\bC}{{\bm C}}
\newcommand{\bTheta}{{\bm \Theta}}
\newcommand{\Inv}{{\bm I}}
\newcommand{\CPT}{{\bC\bP\bT}}
\newcommand{\Psip}{{\psi_+}}
\newcommand{\Psim}{{\psi_-}}
\newcommand{\bra}{{\raise 0.5pt \hbox{$<$}}}
\newcommand{\ket}{{\raise 0.5pt \hbox{$>$}}}
\newcommand{\Ieight}{{1}}
\newcommand{\numeq}{{\;\cong\,}}
\newcommand{\sPsi}{{\psi}}
\newcommand{\bPsi}{{\Psi}}
\newcommand{\etaP}{{e^{i\phi_P}}}
\newcommand{\etaPstar}{{e^{-i\phi_P}}}
\newcommand{\etaT}{{e^{i\phi_T}}}
\newcommand{\etaTstar}{{e^{-i\phi_T}}}
\newcommand{\etaC}{{e^{i\phi_C}}}
\newcommand{\etaCstar}{{e^{-i\phi_C}}}
\newcommand{\Dagger}{{*}}

\begin{document}

\title{
\textbf{On the group generated by $\bC$, $\bP$ and $\bT$:
$\bm{I^2 = T^2 = P^2 = I T P = -1}$,\\ with applications to pseudo-scalar mesons}
}

%

\author{Brian P. Dolan\footnote{email: bdolan@thphys.nuim.ie} \\ \\
  \textit{Department of Theoretical Physics, Maynooth University}\\
  \textit{Main St., Maynooth, Co.~Kildare, Ireland}\\
  and \\ 
  \textit{School of Theoretical Physics}\\
  \textit{Dublin Institute for Advanced Studies}\\
  \textit{10 Burlington Rd., Dublin, Co. Dublin, Ireland}}

\maketitle

\vspace{-12cm}
\rightline{DIAS-STP-20-07}
\vspace{12cm}

\begin{abstract}
We study faithful representations of the discrete Lorentz symmetry operations of parity $\bP$ and
time reversal $\bT$, which involve complex phases when acting on fermions. 
If the phase of $\bP$ is a rational multiple of $\pi$ then $\bP^{2 n}=1$ for some positive integer $n$ and it is shown that, when this is the case, $\bP$ and $\bT$ generate a discrete group, a dicyclic group
(also known as a generalised quaternion group) which  are generalisations of the
dihedral groups familiar from crystallography. 
Charge conjugation $\bC$  
introduces another complex phase and, again assuming rational multiples of $\pi$ for complex phases, 
$\bT \bC$ generates a cyclic group of order $2 m$
for some positive integer $m$.
There is thus a doubly infinite series of possible finite groups labelled by $n$ and $m$. Demanding that $\bC$ commutes with $\bP$ and $\bT$ forces $n=m=2$ and the group generated by $\bP$ and $\bT$ is
uniquely determined to be the quaternion group.

Neutral pseudo-scalar mesons can be simultaneous $\bC$ and $\bP$ eigenstates. 
$\bT$ commutes with $\bP$ and $\bC$ when acting 
on fermion bi-linears so neutral pseudo-scalar mesons can also be $\bT$ eigenstates.
The $\bT$-parity should therefore be experimentally observable and the $\bm{CPT}$ theorem dictates that 
$T= C P$.

\end{abstract}

\newpage

\section{Introduction}

The $CPT$ theorem is a cornerstone of relativistic quantum field theory and
of our understanding of matter at the most fundamental level currently accessible 
to experiment. Parity and time reversal are discrete elements of the Lorentz group 
acting on the Hilbert space of quantum states and they are intimately related to charge conjugation through the $CPT$ theorem.
One can ask what is the discrete group generated by 
the operators $\HatC$, $\HatP$ and $\HatT$?
This question was addressed in \cite{Wigner1} and, choosing the phase of $\widehat P$ so that $\widehat P^2=1$,
three possible groups emerged: ${\mathbb Z}_2\times {\mathbb Z}_2$, ${\mathbb Z}_4\times {\mathbb Z}_2$ and the dihedral group, $D_8$ (in \cite{Wigner2} a similar discussion for the action of $\widehat \Sigma= \widehat C \widehat P$ and $\widehat T$ on physical states gave the same three groups).

When fermionic states are considered there are more possibilities. Racah \cite{Racah} allowed for $\widehat P^2=-1$ on fermionic states and, with $\widehat P^2 =\pm 1$, the list was extended to 8 possible groups in \cite{Shirokov}, 2 of which are specific to spin-1/2 fermions \cite{Socolovsky}.
In this work these possibilities are explored in more detail, the motivation is to construct all possible faithful representations of the improper Lorentz group.

It will be shown that considering $\widehat P$ and $\widehat T$ alone acting on fermions, without reference to $\widehat C$, and assuming only that $\widehat P$ and $\widehat T$ generate a discrete group allows for a doubly infinite series of groups, labelled by two integers $n$ and $m$. Including $\widehat C$, and assuming it commutes with $\widehat P$ and $\widehat T$, reduces this to the two groups in \cite{Socolovsky}, $D_8 \times {\mathbb Z}_2$ and $Q \times {\mathbb Z}_2$, where $Q$ is the quaternion group and ${\mathbb Z}_2$ is generated by $\widehat C$.  The latter group is the only one relevant for Majorana fermions, which are viable candidates for  extensions of the Standard Model that allow for neutrino masses.

The phase $\phi_P$ in the definition of $\widehat P$ does not seem to be physically observable and
Weinberg \cite{Weinberg} has suggested that that it can be removed completely when there is
an internal global $U(1)$ symmetry, such as baryon number or lepton number in the standard model,
in which case $\phi_P$ can be modified by combining $\widehat P$ with a global symmetry phase
transformation to form a modified parity operator $\widehat P'$ satisfying $\widehat P^{\prime\, 2}=1$. The approach here will
not assume any such symmetries associated with underlying dynamics, the properties
of $\widehat P$ will be determined purely in terms of faithful representations of the Lorentz group,
without any reference to a specific model.

Another common argument is that the phase of $\widehat P$ can be ignored because quantum mechanics
only requires projective representations, but this is a dangerous argument.  Phases, such as Aharanov-Bohm and Berry phases,  can have physical effects --- it is safer to work with faithful representations and only project
at the end of the calculation.  This has the added benefit that it provides a check at the end that the physics does not depend on an overall phase.

The strategy therefore is to construct faithful represents of $\widehat P$ and $\widehat T$.
If $\phi_P$  is an irrational multiple of $\pi$ then $\widehat P$
generates a discrete group of infinite order but if the phase is a rational multiple of $\pi$ the group generated by $\widehat P$ is finite.
In this paper all possible finite groups generated by acting on spinors 
with $\widehat C$, $\widehat P$ and $\widehat T$
are classified.

There are some subtleties in discussing these questions when acting on spinors and the associated action of $Sl(2,\mathbb{C}\,)$.  Decompose Fock space into invariant sub-spaces of the Lorentz group and let $\spinor$ denote the sub-space of spinors transforming under the 4-dimensional Dirac representation of the proper Lorentz group in 4-dimensional Minkowski space-time.  Then the definition of $\HatP$, $\HatC$ and $\HatT$ is 
such that $\HatP:\spinor\rightarrow \spinor$ and $\HatC:\spinor\rightarrow \spinor$ but
$\HatT$ complex conjugates $c$-numbers and so maps $\spinor \rightarrow \spinor^\Dagger$.
This means that, while $\spinor$ is an invariant sub-space of $\widehat P$, it is not invariant
under $\widehat T$.

However we can extend $\HatC$, $\HatP$ and $\HatT$ in an obvious way to operators $\bC$, $\bP$ 
and $\bT$ acting on the space
$\spinor \oplus \spinor^\Dagger$ and 
it is then possible to define $\bC$, $\bP$ and $\bT$ so that they do indeed generate a discrete group, and it is non-trivial when acting on Dirac fermions. 

We can summarise the results quite briefly.
Choosing the phase of $\bP$ to be a rational multiple of $\pi$ there is a smallest positive 
integer $n$ for which $\bP^{2 n}=1$ and $\bP$ generates a cyclic group of order $2 n$. The standard definition of $\bT$ implies 
that $\bT^2=-1$ and $\bP$ and $\bT$ then generate a finite group: for $n=1$ it is $\mathbb Z_4 \times \mathbb Z_2$; for $n=2$ it is the quaternion group $Q$ ---  denoting space-time inversion $x\rightarrow -x$ by
$\Inv=\bT\bP$, it is the group defined by
\begin{equation} {\Inv}^2=\bT^2=\bP^2=\Inv \bT\bP=-1.\label{eq:Q}\end{equation} 
For $n \ge 3$ the group generated by $\bP$ and $\bT$ is a generalised quaternion group, $Q_{4 n}$ when $n$ is even 
and Q$_{4 n}\times \mathbb{Z}_2$ when $n$ is odd.\footnote{Generalised quaternion groups Q$_{4 n}$ (also known as dicyclic groups Dic$_n$) are generalisations of the dihedral groups $D_{2n}$, of order $2 n$, familiar from crystallography.
Dihedral groups are generated by rotations through 
$\frac {2 \pi }{n}$ and reflections ${\bm R}$.
Replace ${\bm R}$ (with ${\bm R}^2=1$) by $\bT$ (with $\bT^4=1$) and D$_{2n}$
becomes the generalised quaternion group $Q_{4n}$.}
For $n=2$, Q$_8$ is the quaternion group $Q$, of order eight.

When charge conjugation is included there is another undetermined phase, although
$\bC^2=1$ regardless of this phase. If the phase of $\bC \bT$ is also a rational multiple of $\pi$
then $(\bC \bT)^{2 m}=1$, for some positive integer $m$, and this leads to a doubly infinite set of finite groups, labelled by $n$ and $m$.   This can be reduced to one unique possibility if it is assumed that parity and time reversed transforms of a Majorana spinor are also Majorana, then only $n=m=2$ is allowed and the group consists of two copies of the quaternion group,
$Q \times \mathbb Z_2$, where $\bC$ generates $\mathbb Z_2$.

When $n=2$ the phase of $\widehat P$ is $\pm \frac \pi 2$ (with signature $(1,-1,-1,-1)$, and $\det \widehat P=1$.
A consequence of demanding that two-point correlators $\Delta(x-y) = \{\sPsi(x), \overline \sPsi(y)\}$ vanish at space-like separations is that  $e^{i\phi_P} = \pm i$ for Majorana fermions, \cite{Weinberg}. We can turn this argument around
and say that demanding that $\bC$ commutes with ${\bP}$ implies that $\Delta(x-y)$ vanishes at space-like separations for anti-commuting fields.

Neutral pseudo-scalar mesons, such Kaons and neutral $B$-mesons, can be simultaneous $\bC$ and $\bP$ eigenstates, with
eigenvalues $C P=\pm1$ and $\bT$-violation has been observed in the $B^0-\bar B^0$ system. Although $\bT$ does not commute with $\bP$ on fermions
they do commute on fermion bi-linears.  Neutral pseudo-scalar mesons 
can therefore be simultaneous eigenstates of $\bC$, $\bP$ and $\bT$ 
and the $\CPT$ theorem dictates that their
$\bT$ parity should be the product of $P$ and $C$, so $T=C P=\pm 1$, which
should be experimentally measurable.

The layout of the paper is as follows.
In \S\ref{sec:PT} the algebra of $\bP$ and $\bT$ is shown to be that of $Q_{4n}$ or
$Q_{4 n}\times \mathbb{Z}_2$ for even/odd $n$ respectively.
In \S\ref{sec:CPT} charge conjugation is included and is shown to restrict $n$ to be $2$ if $\bC$ commutes with with $\bP$ and $\bT$. Neutral pseudo-scalar meson systems are discussed in \S\ref{sec:B-mesons}.
Finally the results are summarised and discussed in \S\ref{sec:discussion}. 
Conventions and notation are specified in an appendix but the main text is written so as to be independent of the choice of $\gamma$-matrix conventions.

\section{The group generated by $\bP$ and $\bT$\label{sec:PT}}

Consider a massive spin-$\frac 1 2$ particle with annihilation operators $a_s(\bm p)$ 
and the corresponding anti-particle with creation operators $a^\dagger_{c,s}(\bm p)$, for spin $s=\pm \frac 1 2$.  The field operator is 
\beq \psi(x) = \sum_{s=\pm \frac 1 2} \int \frac{d^3 \bm p}{(2\pi)^3} \bigl\{ u_s({\bm p}) a_s(\bm p) e^{-i p.x} +
 v_s({\bm p}) a^\dagger_{c,s}(\bm p) e^{i p.x}\bigr\},
\label{eq:Psi-Fock}\eeq
where $u_s(\bm p)$ is a positive energy spinor and $v_s(\bm p)$ a negative energy spinor associated with the Clifford algebra
\[ \{ \gamma^\mu,\gamma^\nu \} = 2 \eta^{\mu \nu}.\]
In order to make the analysis completely independent of the conventions
we shall allow for both $\eta_{\mu\nu}=(-1,+1,+1,+1)$ and $\eta_{\mu\nu}=(+1,-1,-1,-1)$ by defining $\varepsilon= \eta^{0 0}=\pm 1$.

Under a Lorentz transformation connected to the identity, $x^\mu \rightarrow \Lambda^\mu{}_\nu x^\nu$,
this transforms to
\beq\psi(x) \rightarrow D(\omega) \psi(\Lambda x), \label{eq:D-omega-psi}\eeq
where $D(\omega)$ is given explicitly in the appendix (\ref{aeq:D-omega}). 
We shall denote the space of all such spinor operators that transform this way by $\spinor$. 

The standard definitions of $\HatP$ and $\HatT$ in second quantisation
are then (see any standard QFT text, {\it e.g.} \cite{Weinberg}, and equations (\ref{aeq:P}) and (\ref{aeq:T}) in the appendix for our conventions for numerical equalities)
\begin{align}
\HatP\sPsi(x) =\, &
\etaP \gamma^0 \sPsi(x_P)
\label{eq:P}\\
\HatT\sPsi(x) =\, & 
\etaT C^* \beta \gamma_5 \gamma^0 \sPsi(x_T)
\label{eq:T}
\end{align}
where $\phi_P$ and $\phi_T$ are complex phase factors, $x_P = (t,-{\bf x})$ denotes the parity transformation on space-time points and $x_T = (-t,{\bf x})$ time-reversal.
$C$ is the charge conjugation matrix and $C^* \beta$ maps $\spinor \rightarrow \spinor^*$.

It is common in quantum field theory textbooks, {\it e.g} \cite{Weinberg}, to write
\[ \HatT\sPsi(x) = -\etaT C \gamma_5 \sPsi(x_T),\]
or even
\[ \HatT\sPsi(x) = \etaT \gamma^1 \gamma^3 \sPsi(x_T),\]
but, while these may be numerically correct in a chosen $\gamma$-matrix representation,  they obscure the Lorentz transformation properties of the spinors.
We wish to write all expressions in a basis independent form and use the notation
\beq \HatT\sPsi(x) =\, 
  \etaT C^* \beta \gamma_5 \gamma^0 \sPsi(x_T)
\numeq  \etaT \gamma^1 \gamma^3 \sPsi(x_T),\label{eq:T-numeq}\eeq 
where notation $=$ is reserved for equations involving matrix multiplication that 
preserve the spinor index structure, it might be called a spinorial equality, while $\numeq$ indicates equations that do not preserve the spinor index structure, but are nevertheless numerically 
correct in a specifically chosen basis (more details are given in the appendix).  If one is confident the index structure is correct $\numeq$ is an extremely useful calculational tool, but it should be used with care when discussing the Lorentz transformation properties of spinors.\footnote{The spinorial equality in equation (\ref{eq:T}) is obtained by carefully tracking the spinor index structure in the derivation of $\widehat T$ in the chiral representation, the numerical equality is probably the more familiar one, with signature $(+,-,-,-)$.} 

The square of the parity operator is proportional to the identity,  $\hat P^2 = e^{2i\phi_P}(\gamma^0)^2 =  \varepsilon e^{2i\phi_P}$, where
$\varepsilon= \eta^{0 0}$. Of course the sign of $\eta^{0 0}$ can
be absorbed into the phase $\phi_P$, so it cannot matter which sign we choose for the metric,
but we shall keep $\varepsilon$ explicit in order to follow through the mathematical consequences
of the different possible conventions.

  If $\sPsi\in\spinor$ and $\sPsi\,'\in\spinor$ are two Dirac 
spinors then
$\sPsi^\Dagger \in \spinor^\Dagger$ and
\[\overline {\sPsi}=  \sPsi^\dagger \beta\]
 is in the dual vector space to $\spinor$, which will be denoted by $\spinor^D$. For ordinary complex vector spaces this would mean that $\overline{\sPsi} \sPsi\,' \in {\mathbb C}$, but for spinors
\[  \overline{\sPsi} \sPsi\,' = \sPsi^\dagger \beta \sPsi\,'\]
is a Lorentz invariant Fock space operator in second quantisation. $\beta$ is an hermitian metric on $\spinor$, 
$\beta:\spinor \rightarrow \bigl(\spinor^D\bigr)^\dagger$ and $\beta^\dagger = \beta$  while 
$C:\bigl(\spinor^D\bigr)^T \rightarrow \spinor$.
The important properties of $\beta$ and $C$  are
\[ \beta \gamma^\mu = \varepsilon \bigl(\gamma^\mu\bigr)^\dagger \beta, 
\qquad \gamma^\mu C = - C \bigr(\gamma^\mu\bigr)^T, 
\qquad C \beta^* \bigl(\gamma^\mu\bigr)^* = - \varepsilon \gamma^\mu C \beta^*.\] 
which is all that will be used in  the ensuing analysis.

Let ${\bm 1}$ denote the identity matrix acting on
 $\spinor$ and $\bar {\bm 1}$ the identity matrix on $\spinor^\Dagger$ then,
using a chiral basis in which 
\[\beta = \bpm 0 & \bar {\bf 1} \\ {\bm 1} & 0 \epm,\]
gives $\beta\numeq \beta^\tr$ and $\beta^2\numeq 1$, 
but $\beta\ne\beta^\tr$ and $\beta^2\ne 1$.  However $\beta^\dagger=\beta$ 
is correct.  With the conventions in appendix \ref{sec:conventions}\footnote{From now in $\numeq$ will
refer to the  conventions in the chiral basis of appendix \ref{sec:conventions}, with
$\varepsilon=+1$, but $\varepsilon$ will be kept explicit in spinorial equalities.}
\[\gamma^0\numeq \bpm 0 & {\bf 1} \\ {\bm 1} & 0 \epm\]
so $\beta\numeq\gamma^0$ but $\beta\ne \gamma^0$.
Similarly $\beta\gamma^0\numeq 1$ and in our conventions $C$ is real, $C^*\numeq C$,
also $\beta\gamma_5\gamma^0 \numeq -\gamma_5\beta\gamma^0\numeq-\gamma_5$ hence equation (\ref{eq:T-numeq}). Spinorial equations involving $=$ are valid in any consistent
set of conventions for the $\gamma$-matrices, numerical equations involving $\numeq$ can be convention dependent, though they are always the same in unitarily equivalent representations of the $\gamma$-matrices once the sign of $\eta^{0 0}$ is chosen, see {\it e.g.} appendix A-2 of \cite{I+Z}.

The structure of (\ref{eq:P}) and (\ref{eq:T})  is perhaps a little clearer if
they are re-written as
\begin{align}
\HatP\sPsi(x) =\, &
i\etaP \gamma_5 \gamma^{1 2 3} \sPsi(x_P)   \\
\HatT\sPsi(x) =\, & 
\etaT C^* \beta \gamma_5 \gamma^0 \sPsi(x_T),
\end{align}
the index structure is then more logical and $C^* \beta$ is the matrix
that maps $D(\omega) \rightarrow D^*(\omega)$.
As described in the appendix $\Hat P:\spinor \rightarrow \spinor$ and 
$\Hat T:\spinor \rightarrow \spinor^\Dagger$.

It is straightforward to define
\begin{align}
\HatP^\Dagger\sPsi^\Dagger(x) = \{ \widehat P \psi(x) \}^*=\, &
\etaPstar \bigl(\gamma^0\bigr)^* \sPsi^\Dagger(x_P)
\label{eq:P*}\\
\HatT^\Dagger\sPsi^\Dagger(x) = \{ \widehat T \psi(x) \}^* =\, & 
\etaTstar C \beta^* \gamma_5^* \bigl(\gamma^0\bigr)^* \sPsi^\Dagger(x_T)
\numeq -\etaTstar C \gamma_5 \sPsi^\Dagger(x_T).\label{eq:T*}
\end{align}
Then $\HatP^\Dagger:\spinor^\Dagger \rightarrow \spinor^\Dagger$ and  
$\HatT^\Dagger:\spinor^\Dagger \rightarrow \spinor$, with $\HatT^\Dagger=-\HatT^{-1}$.
$\spinor$ is a sub-space of the full Hilbert space that is
an invariant sub-space under $\widehat P$ but it is not an invariant
sub-space under $\widehat T$.

In the chiral basis in the appendix 
\begin{align}
\HatP\sPsi(x) &\numeq \etaP \beta \sPsi(x_P) \numeq \etaP \bpm 0& 1 \\ 1 & 0 \epm \sPsi(x_P)
\label{eq:nP}\\
\HatP^\Dagger\sPsi^\Dagger(x) &\numeq e^{-i \phi_P} \beta \sPsi^\Dagger(x_P)
\numeq e^{-i \phi_P} \bpm 0 & 1 \\ 1 & 0 \epm \sPsi^*(x_P)
\label{eq:nP*}\\
\HatT\sPsi(x) & 
\numeq \etaT \bpm \epsilon & 0 \\ 0 & \epsilon \epm \sPsi(x_T)
\label{eq:nT}\\
\HatT^\Dagger\sPsi^\Dagger(x) & 
\numeq \etaTstar \bpm \epsilon & 0 \\ 0 & \epsilon \epm \sPsi^\Dagger(x_T)\label{eq:nT*}\end{align}
where $\epsilon \numeq \bpm 0 & 1 \\ -1 & 0 \epm$.

The strategy is now to double spinor space to $\spinor \oplus \spinor^\Dagger$ and construct 8 component spinors
\[ {\bPsi}(x) = \bpm \sPsi(x) \\ \sPsi^\Dagger(x) \epm. \]
This is not a general element of $\spinor \oplus \spinor^\Dagger$, it is restricted so that the lower component is the complex conjugate of the upper component, and the space of such restricted spinors will be denoted by $\Spinor$.  
This idea of doubling the space of states was introduced in \cite{Wigner2}, though in the present context it is really more of a grading
of $\Spinor$, into one particle states that transform under the Lorentz group under $D(\omega)$ and those that transform under $D^*(\omega)$,
rather than a doubling of the space. We really want the action of $\hat P$ and $\hat T$ on the full Hilbert space and neither $\spinor$ nor $\Spinor$ is the
full Hilbert space, with consists of multi-particle states. But it is sufficient to determine the action of $\hat P$ and $\hat T$ on $\Spinor$ in order to determine their action on the
full Hilbert space.  $\spinor $ is an invariant subspace of $\hat P$ but not
of $\hat T$, it is necessary to go to $\spinor \oplus \spinor^*$ to get
an invariant sub-space of $\hat T$ if we wish to evaluate the full group
generated by $\hat P$ and $\hat T$.

On $\Spinor$ the action of  parity and time-reversal on the spinor indices of $\bPsi(x)$ is represented 
as ordinary multiplication by the $8\times 8$ matrices
\begin{align} 
\bP & = \bpm \etaP \gamma^0 & 0 \\ 0& \etaPstar \bigl(\gamma^0\bigr)^*\epm \label{eq:bP-def}\\
\bT &=\bpm 0 & \etaTstar C \beta^* \gamma_5^* \bigl(\gamma^0\bigr)^* \\
\etaT C^* \beta \gamma_5 \gamma^0& 0 \epm.\label{eq:bT-def}
\end{align}
We are now in a position to determine the group generated
by $\bP$ and $\bT$ using straightforward matrix multiplication. 
Using the relation (\ref{eqa:bCbC})
\[C^* \beta \gamma^\mu = -\varepsilon(\gamma^\mu)^* C^* \beta, \] 
it is immediate from (\ref{eq:bP-def}) and (\ref{eq:bT-def}) that\footnote{These matrices are written in $4\times 4$ block form.
To avoid otherwise cumbersome notation both the $4\times 4$ and the  $8\times 8$ identity matrices
are just signified by $1$.}
\begin{align} 
\bP^{ 2k} &= \varepsilon^k\bpm e^{2 k i \phi_P}  & 0 \\ 0 & e^{-2 k i \phi_P} \epm\label{eq:P^2k}\\
\bP^{ 2k+1} &= \varepsilon^k\bpm e^{(2 k+1) i \phi_P}   \gamma^0 & 0 \\ 0 & e^{-(2 k+1) i \phi_P}  
(\gamma^0)^* \epm
\label{eq:P^2k+1}\\
\bT^2 & = -{\Ieight}\label{eq:Tsquared}\\
 (\bP \bT)^2 &=-1, \qquad \bP \bT \bP = \bT. \label{eq:PTP=T}
\end{align}  
The phase $\phi_P$ and signature $\varepsilon$ drop out of
(\ref{eq:Tsquared}) and (\ref{eq:PTP=T})
and we immediately note that
\[\bP \bT = \bT \bP^{-1}\] 
so $\bP$ and $\bT$ only commute if $\bP^2=1$.

If $\phi_P$ is an irrational multiple of $\pi$ the algebra never closes and the group generated by $\bP$ is of infinite order.  But if $\phi_P$ is a rational multiple of $\pi$ then there exits a smallest positive integer $n$ for which 
\beq\bP^{2 n}={\Ieight}\label{eq:P^n} \eeq
and $\bP$ generates the cyclic group ${\mathbb Z}_{2 n}$.

To understand the structure in more detail we analyse $\varepsilon=+1$
and $\varepsilon=-1$ separately.

\begin{itemize}

\item $\pmb{\varepsilon=+1:}$ First consider specific cases with small $n$:

\begin{itemize}
\item[$\bullet$] $n=1$, $\phi_P=\pm\pi$: $\bP^{-1}=\bP$ and (\ref{eq:PTP=T}) implies that $[\bT,\bP]=0,$
so $\bT$ and $\bP$ generate the abelian group ${\mathbb Z}_4\times {\mathbb Z}_2$.

\item[$\bullet$] $n= 2$, $\phi_P=\pm\frac \pi 2$: $\bP^2=-{\Ieight}$ 
and (\ref{eq:PTP=T})
 implies that $\bT \bP = -\bP \bT$.  Let $\Inv=\bT\bP$ denote space-time inversion,  sending space-time points $x\rightarrow -x$, then 
\beq \Inv^2 = \bT^2 = \bP^2 =\Inv\bT\bP=-{\Ieight}\label{eq:Q-algebra}\eeq 
generates the quaternion group, $Q$.\footnote{We are working with the complex  Clifford algebra. The appearance of $Q$ here does not seem to have any obvious relation to the appearance of quaternions in the real Clifford algebra of $2\times 2$ matrices of quaternions.}

\item[$\bullet$] $n=3$,  $\phi_P=\pm\frac \pi 3$: in this case (\ref{eq:P^2k+1}) implies 
$\bP^3=-\bpm \gamma^0 & 0 \\ 0 & (\gamma^0)^* \epm $. 
Let
\[ \bV = \bpm \gamma^0 & 0 \\ 0 & (\gamma^0)^* \epm 
\qquad \hbox{and} \qquad \bU =  \bpm e^{\pm i \pi/3} {\bm 1}& 0 \\ 0 & e^{\mp i \pi /3} \bar {\bm 1}\epm. \]
then $\bU^3=-1$ and $[\bP,\bm \bV]=[\bT,\bm \bV]=0$.
From (\ref{eq:PTP=T}) 
\[\bT^{-1} \bU\, \bT = \bU^{-1}\]
and we have a finite group with presentation
\beq
<\bU,\bT\, |\, \bU^6=1, \bU^3 =\bT^2, \bT^{-1} \bU \bT = \bU^{-1}>,\label{eq:Dic_3}
\eeq 
which is the generalised quaternion  group Q$_{12}$ of order 12. Including $\bm \bV$ adds
a trivial $\mathbb Z_2$ making the full group Q$_{12} \times \mathbb{Z}_2$. The strategy is to start with $\bP$ and $\bT$, use these to define
$\bV =\bP^3 \bT^2$ and $\bU = \bP \bV$, then $\bT$ and  $\bU$
(with the presentation (\ref{eq:Dic_3})),
together with  $\bV$, give the full group. 
\end{itemize}
The cases $n=2$ and $n=3$ generalise to two infinite families:
\begin{itemize}
\item[$\bullet$] even $n$:
$ \bP^n = \bT^2 = -{\Ieight} $
and together $\bP$ and $\bT$ generate the generalised quaternion group Q$_{4 n}$ of order $4n$ with the presentation
\beq
\mbox{Q}_{n4}=<\bP,\bT\, |\, \bP^{2 n}={\Ieight}, \,\bP^n=\bT^2, \,\bT^{-1} \bP \bT = \bP^{-1}>.
\eeq 
Q$_8$ is the quaternion algebra.
\item[$\bullet$] odd $n>1$: with
  \beq \bV = \bpm \gamma^0 & 0 \\ 0 & (\gamma^0)^*\epm,\label{eq:V_+}\eeq
  $\bP = \bU\, \bV$,
$\bU$ and $\bT$ generate the generalised quaternion  group Q$_{4 n}$ with the presentation
\beq
\mbox{Q}_{4n} = 
<\bU,\bT\, |\, \bU^{2 n}={\Ieight}, \bU^n=\bT^2, \bT^{-1} \bU \bT = \bU^{-1}>
\eeq
and, including $\bV=\bP^n \bT^2$, the full group is  Q$_{4 n} \times \mathbb{Z}_2$.
\end{itemize} 

\item $\pmb{\varepsilon =-1}:$ the analysis is very similar to $\varepsilon=+1$, but with slight modifications due to the fact that $\bT$ anti-commutes
  with $\bpm \gamma^0 & 0 \\ 0 & (\gamma^0)^* \epm$ for this choice.
\begin{itemize}
\item[$\bullet$] $n=1$, $\phi_P=\pm\frac{\pi}{2}$: $\bP^{-1}=\bP$ and (\ref{eq:PTP=T}) implies that $\bT\bP=\bP\bT.$
  Again $\bT$ and $\bP$ simply generate abelian the
  group ${\mathbb Z}_4 \times {\mathbb Z}_2$.

\item[$\bullet$] $n= 2$, $\phi_P=0,\pi $: so $\bP^2=\bT^2=-1$ 
  and $\bT \bP = -\bP \bT$. As for $\varepsilon=+1$,  $\Inv =\bT\bP$ and $\Inv^2=-1$,
  and the  group is again the quartenion group $Q$.

\item[$\bullet$] $n=3$,  $\phi_P=\pm\frac \pi 6$: then
\[\bU:=-\bP^2 = \bpm e^{\pm \pi i/3}{\bm 1} & 0 \\ 0 &  e^{\mp \pi i/3} \bar {\bm 1}\epm\]
  generates ${\mathbb Z}_6$ and $\bU^3=-1$.
  Also
  \beq \bV:=\bP^3= \mp \bpm i\gamma^0  & 0 \\ 0 & -i (\gamma^0)^*\epm\label{eq:V_-}\eeq
  generates ${\mathbb Z}_2$.  This gives $[\bT,\bV]=0$ and, trivially, $[\bU,\bV]=0$ so $\bV$ just generates a ${\mathbb Z}_2$
  factor and we only need to worry about $\bT$ and $\bU$.
  We have  $\bT \bU = \bU^{-1} \bT$ and so $\bT$ complex conjugates $\bU$.
  Let $\bW = \bU\bT$ then
  \[<\bU,\bT,\bW |\,\bU^6=1,\bU^3 = \bT^2= \bW^2 = \bU \bT \bW> \]
  is another presentation of $\mbox{Q}_{12}$. As for $\varepsilon = +1$ the full
  group is $\mbox{Q}_{12} \times {\mathbb Z}_2$.
  
\item[$\bullet$] $n=4$, $\phi_P = \pm \frac{\pi}{4}$. $\bP^4=-1$ and, from (\ref{eq:PTP=T}),
  $\bT^{-1} \bP \bT= \bP^{-1}$.
 We have the  group with presentation
  \beq
<\bP,\bT\, |\, \bP^8={\Ieight}, \,\bP^4 = \bT^2, \bT^{-1} \bP \bT= \bP^{-1}>.
\eeq 
This is the generalised quaternion group Q$_{16}$.

\end{itemize}

\item[$\bullet$] The cases $n=3$ and $n=4$ generalise to two infinite families

  \begin{itemize}

  \item[$\bullet$] even $n\ge 4$: $\bP^ n=-1$ and the group is again Q$_{4 n}$,
    \[<\bP,\bT\, |\, \bP^{2 n} ={\Ieight}, \,\bP^n = \bT^2, \bT^{-1} \bP \bT= \bP^{-1}>.\]
   
  \item[$\bullet$] odd $n\ge 3$: $\phi_P = \pm \frac{\pi}{2 n}$, define
    $\bU:=-\bP^2 = \bpm e^{\pm \pi i/n} {\bm 1}& 0 \\ 0 &  e^{\mp \pi i/n} \bar {\bm 1}\epm$
    which  generates ${\mathbb Z}_{2 n}$, with $\bU^n=-1$.

    Let
  $\bV:=\bP^n= \mp \bpm i\gamma^0 & 0 \\ 0 & -i(\gamma^0)^* \epm$ generate ${\mathbb Z}_2$.
  
  With $[\bT,\bV]=[\bU,\bV]=0$, $\bT \bU = \bU^{-1} \bT$ and $\bW = \bU \bT$ we have
  \[<\bU,\bT,\bW |\,\bU^{2 n}=1,\bU^n = \bT^2= \bW^2 = \bU \bT \bW> \]
and the full group is Q$_{4 n} \times {\mathbb Z}_2$.

\end{itemize} 

The group structure is the same with either choice of sign for the signature:
${\mathbb Z}_4 \times {\mathbb Z}_2$ for $n=1$, 
Q$_{4 n}$ for even $n$ and Q$_{4 n} \times {\mathbb Z}_2$ for odd $n\ge 3$.

\end{itemize}

This completes the classification of all possible finite groups generated by $\bP$ and $\bT$ acting on Dirac spinors in $\Spinor$. We now go on to discuss the unitary properties of $\bP$ and $\bT$. 
Numerically $\bP$ and $\bT$ are unitary matrices 
\[ \bP^\dagger \numeq \bP^{-1}, \qquad \bT^\dagger \numeq \bT^{-1},\]
but these are not spinorial equalities.
The definition of a unitary operator requires defining a bi-linear form on $\Spinor$ and there is a
natural  hermitian Lorentz invariant metric,
\beq{\bm \beta}_+ = \frac 1 2 \bpm \beta & 0 \\ 0 & \beta^* \epm.\label{eq:beta_+}\eeq
For two spinors ${\bPsi} \in \Spinor$ and $ {\bPsi}' \in \Spinor$
\[\widetilde  {\bPsi} \bPsi':= \bPsi^\dagger {\bm \beta}_+ \bPsi' 
= \frac 1 2 \bigl(\sPsi^\dagger \beta \sPsi\,' + \sPsi^\tr\beta^* \sPsi\,'^\Dagger\bigr)=
\frac 1 2 \bigl(\sPsi^\dagger \beta \sPsi\,' - \sPsi\,'^\dagger \beta \sPsi\bigr)
=\frac 1 2 \bigl(\overline \psi \psi' - \overline{\psi'} \psi\bigr)
\]
is Lorentz invariant (the minus sign is due to anti-commutativity of fermions). 

This gives
\beq \widetilde {(\bP \bPsi)} (\bP \bPsi') =  \widetilde {(\bT \bPsi)} (\bT \bPsi') =
\widetilde \bPsi \bPsi'\label{eq:PTbeta_+}\eeq 
and both $\bP$ and $\bT$ are unitary with this bi-linear form.\footnote{To avoid cumbersome  notation the space-time argument
of $\bPsi$ is omitted, it should be clear from the context how $x$ is affected. In any case he metric on the
full Hilbert space of position dependent spinors includes $\int d^4 x$ and the argument of $\bPsi$ disappears from the inner product when the integral is evaluated.} 
Unfortunately 
\[ \widetilde {\bPsi} \bPsi=0\]
and this is of little use in constructing a Lagrangian.
An alternative is to use
\beq{\bm \beta}_- = \frac 1 2 \bpm \beta & 0 \\ 0 & -\beta^* \epm\label{eq:beta_-}\eeq
and
\begin{align}
\overline{\bPsi} \bPsi':= \bPsi^\dagger {\bm \beta}_- \bPsi' 
  &= \frac 1 2 \bigl(\sPsi^\dagger \beta \sPsi\,' - \sPsi^\tr\beta^* \sPsi\,'^\Dagger\bigr)
  =\frac 1 2 \bigl(\sPsi^\dagger \beta \sPsi\,' + \sPsi\,'^\dagger \beta \sPsi\bigr)\nonumber \\
&=\frac 1 2 \bigl(\overline \psi \psi' + \overline{\psi'} \psi\bigr).
\label{eq:PsibarPsi}
\end{align}
This renders $\bP$ unitary but this bi-linear is not invariant under $\bT$
\beq  \overline {(\bP \bPsi)} (\bP \bPsi') =  -\overline {(\bT \bPsi)} (\bT \bPsi') =
\overline \bPsi \bPsi'\label{eq:Tminus}
\eeq
giving\footnote{Equations (\ref{eq:PTbeta_+}) and (\ref{eq:Tminus}) are most easily proven using
\begin{align}
\bP^\dagger {\bm \beta}_\pm & = {\bm \beta}_\pm \bP^{-1}\\
 \bT^\dagger {\bm \beta}_\pm &= \mp   {\bm \beta}_\pm \bT.
\end{align}
}
\beq  \overline {(\bT \bPsi)} (\bT \bPsi)
= - \overline \bPsi \bPsi.\label{eq:T-}\eeq

The standard action of $\widehat T$ on 4-component spinors in $\spinor$
gives
\[\widehat T ( \overline \psi \psi)=
  \overline{(\widehat T \psi)} \beta^* (\widehat T \psi)=\overline \psi \psi\]
and the reason for this difference in sign is the anti-unitary nature
of $\widehat T$, which interchanges $\beta$ and $\beta^*$.
Alternatively the sign change can be accounted for by demanding that $\widehat \bT$ reverses the order of the spinors (as observed in \cite{Bell} this is equivalent to $\beta\rightarrow -\beta$ in fermion bi-linears).  
This property of time reversal can be accommodated in $\Spinor$ by defining
\beq \widehat \bT (\overline \bPsi \bPsi') :=
-\overline {(\bT \bPsi)} (\bT \bPsi')
=\overline {\bPsi'} \bPsi,\label{eq:T-hat}
\eeq
while
\[ \widehat \bP (\overline \bPsi \bPsi') := \overline {(\bP \bPsi)} (\bP \bPsi') = \overline \bPsi \bPsi',\]
remains unchanged. 

On $\spinor$ time reversal $\widehat T$ is an anti-linear operator: if $\psi$ is multiplied by a phase, $\psi \rightarrow e^{i \alpha}\psi$, then $\widehat T (e^{i \alpha} \psi)=e^{-i\alpha} \widehat T \psi$.
In ${\Spinor}$ this is represented by
\[ \bT \bpm e^{i \alpha} \psi \\ e^{-i \alpha} \psi^* \epm
= \bT   \bpm e^{i \alpha} & 0  \\ 0 & e^{-i \alpha}   \epm \bPsi
= \bpm e^{-i \alpha}  & 0  \\ 0 & e^{i \alpha}  \epm \bT \bPsi\]
and this is simply matrix multiplication, it does not involve any explicit complex conjugation.
Define the $8\times 8$ matrix, in $4\times 4$ block form,
\[ {\cal I}:=\bpm i  & 0 \\ 0 & -i  \epm\]
then $\Psi \rightarrow e^{{\cal I}\alpha} \Psi$ and matrix multiplication 
gives
\[ \bT {\cal I} = - {\cal I} \bT, \]
with no need to complex conjugate $c$-numbers explicitly.
At the same time it is trivial that
\[ \bP {\cal I} = {\cal I} \bP,\]
thus space-time inversion anti-commutes with ${\cal I}$
\[ \Inv {\cal I} = -{\cal I} \Inv.\]
Note that multiplication by ${\cal I}$ interchanges $\bm {\beta}_+$ and $\bm {\beta}_-$ in that
\[ \overline \Psi {\cal I} \Psi' = i\widetilde \Psi \Psi'.\]
There is therefore no need to introduce anti-linear operators in $\Spinor$. The only difference between $\widehat \bT$ and $\widehat \bP$
in this regard is that $\widehat \bT$ reverses the order of spinors whereas $\widehat \bP$ preserves the order.

Now consider the reducible representation of the Clifford algebra 
\[ \bm\Gamma^\mu = \bpm \gamma^\mu & 0 \\ 0 & \varepsilon \bigl(\gamma^\mu\bigr)^*
\epm,
\]
giving
\[ \{ \bm \Gamma^\mu,\bm \Gamma^\nu\} = 2 \eta^{\mu\nu},\]
in terms of which 
\[\overline \bPsi\, {\bm\Gamma}^\mu \bPsi' 
= \frac 1 2 \bigl( \overline \sPsi \gamma^\mu \sPsi\,' + \overline { \sPsi\,' } \gamma^\mu \sPsi\bigr),
\quad
\widetilde \bPsi\, \Gamma^\mu \bPsi' =
 \frac 1 2 \bigl( \overline  \sPsi \gamma^\mu \sPsi\,' - \overline { \sPsi\,' } \gamma^\mu \sPsi\bigr).\]
The actions of $\bP$ and $\bT$ on these are
\begin{align} 
\widehat \bP \bigl( \overline {{\bPsi}}\, \bm \Gamma^\mu {\bPsi'}\bigr)
&=  {\cal P}^\mu{}_\nu  (\overline {\bPsi}\bm \Gamma^\nu\bPsi'),
\qquad
\widehat \bP \bigl( \widetilde {{\bPsi}}\, \bm \Gamma^\mu {\bPsi'}\bigr)
=  {\cal P}^\mu{}_\nu  (\widetilde {\bPsi}\bm \Gamma^\nu\bPsi')\\
 \widehat \bT \bigl(\overline {{\bPsi}}\, \bm \Gamma^\mu  {\bPsi'}\bigr)
&= {\cal P}^\mu{}_\nu \bigl(\overline {\bPsi'}  \bm \Gamma^\nu\bPsi\bigr),
\hskip 17pt 
 \widehat \bT \bigl(\widetilde {{\bPsi}}\, \bm \Gamma^\mu  {\bPsi'}\bigr)
= - {\cal P}^\mu{}_\nu \bigl(\widetilde {\bPsi'}  \bm \Gamma^\nu\bPsi\bigr) \label{eq:TminusGamma5}
\end{align}
where $ {\cal P}=
  {\scriptstyle\bpm 1 & 0 & 0 & 0\\ 
0 & -1 & 0 & 0 \\ 
0 & 0 & -1 & 0 \\ 
0 & 0 & 0 & -1 \epm}$
implements the parity operation on space-time points, $x\rightarrow x_P$.

With
$ \bm \Gamma_5 =  \bpm \gamma_5 & 0 \\ 0 & \gamma_5^* \epm$ 
\[\overline \bPsi\, \Gamma_5 \bPsi' = \frac 1 2 \bigl(\overline \sPsi \gamma_5 \sPsi\,' - \overline {\sPsi\,'} \gamma_5 \sPsi\bigr),
\qquad 
\widetilde \bPsi \Gamma_5 \bPsi' = \frac 1 2 \bigl(\overline \sPsi \gamma_5\sPsi\,' + \overline {\sPsi\,'} \gamma_5\sPsi\bigr)
\] 
and 
\begin{align}
\widehat \bP (\overline \bPsi \Gamma_5 \bPsi')=\overline  {\big(\bP \bPsi\bigr)}\, \bm \Gamma_5 \bigl({\bP \bPsi'}\bigr)
&=-\overline \bPsi \Gamma_5 \bPsi'\\ 
\widehat \bT \bigl(\overline  {\bPsi} \, \bm \Gamma_5  \bPsi\,'\bigr)
=-\overline  {\big(\bT \bPsi\bigr)}\, \bm \Gamma_5 \bigl({\bT \bPsi'}\bigr)
&=\overline {\bPsi'} \Gamma_5 \bPsi\\
\widehat \bP (\widetilde \bPsi \Gamma_5 \bPsi')=\widetilde  {\big(\bP \bPsi\bigr)}\, \bm \Gamma_5 \bigl({\bP \bPsi'}\bigr)
&=-\widetilde \bPsi \Gamma_5 \bPsi'\\ 
\widehat \bT \bigl(\widetilde  {\bPsi} \, \bm \Gamma_5  \bPsi\,'\bigr)
=-\widetilde  {\big(\bT \bPsi\bigr)}\, \bm \Gamma_5 \bigl({\bT \bPsi'}\bigr)
&=-\widetilde {\bPsi'} \Gamma_5 \bPsi. \label{eq:Tminus5}
\end{align} 

\section{The groups generated by $\bC$, $\bP$ and $\bT$ \label{sec:CPT}}

The charge conjugation operator is defined as 
\beq
\HatC\sPsi(x) = \etaC C \beta^\tr \sPsi^\Dagger(x)\numeq \etaC C \beta \sPsi^\Dagger(x)\label{eq:C}
\eeq
with
\[ C=\bpm \epsilon^{-1} & 0 \\ 0 & \bar \epsilon\,^{-1} \epm, \]
$\epsilon \numeq \bpm 0 & 1 \\ -1 & 0 \epm$ and $\phi_C$ another complex phase.
The charge conjugation matrix $C$ maps $(\spinor^D)^T \rightarrow \spinor$ and is defined 
by demanding that $\gamma^\mu C = -(\gamma^\mu)^T C$.

This leads to the natural definition of the $8\times 8$ matrix
\beq 
\bC = \begin{pmatrix}  
0 & \etaC C \beta^* \\ 
\etaCstar C^* \beta & 0 
\end{pmatrix} 
\numeq 
\bpm 0 & \etaC C \beta \\ 
\etaCstar C \beta &0 \epm
\label{eq:bC-def}
\eeq
acting on $\Spinor$ and, using (\ref{aeq:Cbeta-relation}),
\[  \bC^2={\Ieight}.\]
From (\ref{eq:bP-def}) and (\ref{eq:bT-def})
\begin{align}
(\bC \bP)^2&= -1\\
(\bC \bT)^2 &= -\varepsilon\bpm e^{2 i (\phi_T + \phi_C)} & 0 \\ 0 & e^{-2 i (\phi_T + \phi_C)}  \epm
=-\varepsilon \,e^{2 {\cal I}(\phi_T + \phi_C)}.
\label{eq:CPTgroup-relations}
\end{align}
If $\phi_T + \phi_C$ is an irrational multiple of $\pi$ 
the algebra does not close and the group is of infinite order.
But if $\phi_T + \phi_C$  
is a rational multiple of $\pi$ then there is a smallest positive integer $p$ for which
$e^{i p (\phi_T + \phi_P)}=1$
 and (\ref{eq:CPTgroup-relations}) then implies that 
 \begin{align}
   \varepsilon  =& 1 & \varepsilon =-&1 & \nonumber\\
  & \kern -40pt\rule{60pt}{1pt}& &  \kern -40pt \rule{60pt}{1pt}& \nonumber\\
   (\bC \bT)^{4 p} &=1, & (\bC \bT)^{2 p} &=1, & \mbox{if} \  p \ \mbox{is odd,}\hskip 25pt \  \nonumber\\
(\bC \bT)^{2 p}  & = 1, & (\bC \bT)^{p}  & = 1, & \mbox{if} \  p=2 \mod 4,  \label{eq:CT^p=1} \\
(\bC \bT)^p  & = 1, & (\bC \bT)^p  & = 1,  & \mbox{if} \  p=0 \mod 8,\nonumber\\
   (\bC \bT)^{\frac p 2}  & = 1, & (\bC \bT)^p  & = 1, & \mbox{if} \  p=4 \mod 8,\nonumber
 \end{align}
and these are all the lowest powers of $\bC\bT$ that give 1 on the right hand side.
In general $\bC$ need not commute with either $\bP$ or $\bT$,
\begin{align}
\bC \bP &=-\bP^{-1} \bC
\qquad \Rightarrow \qquad \bP \bC \bP = -\bC\\
\bC \bT &=\varepsilon e^{\frac{2 \pi {\cal I}}{p} } \bT \bC.
\label{eq:[C,T]}
\end{align}
There is a doubly infinite series of possible finite groups generated by 
$\bP$, $\bT$ and $\bC$, characterised by the two integers $n$ and $m$: 
\begin{itemize}
\item $n$ even, 
\[
<\bP,\bT,\bC: \bP^{2 n}=\bT^4=\bC^2= (\bP \bT)^4=(\bC \bP)^4=(\bC \bT)^{2 m}=1>\]
\item $n$ odd, 
\begin{align*}
  <\bU,\bV,\bT,\bC:\, & \bU^{2 n}=\bT^4=\bV^2=\bC^2
                        = \bigl(\bU \bV \bigr)^{2 n}= (\bU \bT)^4\\
&= \bigl(\bC \bU\bigr)^2=
\bigl(\bV \bT\bigr)^2=\bigl(\bC \bV\bigr)^2= (\bC \bT)^{2 m}= 1>,
\end{align*}
\end{itemize}
where $m$ can be read off from equations (\ref{eq:CT^p=1}) for either choice of $\varepsilon$ and any given $p$, with $\bV$ given by (\ref{eq:V_+}) for $\varepsilon=+1$ and (\ref{eq:V_-}) for $\varepsilon=-1$.

This doubly infinite series can be reduced to one unique group if we  assume that $\bC$ commutes with $\bP$ and $\bT$.
Let $\bPsi$ be a Majorana spinor,
$\bC \bPsi(x) = \pm \bPsi(x)$,
and assume that the parity reversed state of $\bPsi$ is also Majorana with the same $C$-parity, it seems reasonable to assume that a parity transformation does not change the Majorana property of a spinor.
Then\footnote{If $\bC \bPsi = - \bPsi$, then let $\bPsi' = {\cal I} \bPsi$ and $\bC \bPsi' = \bPsi'$.}
\[ \bC \bP \bPsi(x) = \bP \bPsi(x) = \bP \bC \bPsi(x)
\]
and
\[ [\bC,\bP]=0\]
which requires $e^{i\phi_P}=\pm i$, so $n=2$.
The possibility of $\bP^2=-1$ for fermions was introduced by Racah \cite{Racah} and
it is shown in \cite{Weinberg} that it is necessary for Majorana spinors, but this latter argument
requires the assumption that 2-point functions must vanish at space-like separations. We see here that
it can be deduced purely algebraically from the assumption that $[\bC,\bP]=0$.

The same reasoning can be applied to time reversal, assume that the time reversed state of a Majorana spinor $\bPsi$ is also Majorana,
\[ \bC \bT \bPsi(x)  = \bT \bPsi(x) = \bT \bC \bPsi(x).
\]
Then
\[ [\bC,\bT]=0\]
and $\varepsilon e^{\frac{4\pi{\cal I}}{p}}=1$ which requires $p=1$ or $2$ for $\varepsilon=+1$ and $p=4$ for $\varepsilon=-1$, in all three cases $\bigl(\bC \bT\bigr)^4=1$ and $m=2$.
The group is then
\beq
<\bP,\bT,\bC: \bP^2=\bT^2=(\bT\bP)^2=-1, \bC^2=1,(\bP \bC)^2=(\bT \bC)^2=-1>,
\label{eq:Q-alebra}\eeq
which is a different presentation of $Q\times {\mathbb Z}_2$: the group generated by $\bP$ and $\bT$ is the quaternion group (\ref{eq:Q-algebra}) and the full group is two copies of the quaternion group, the $\mathbb Z_2$ being generated by $\bC$.

A general Dirac spinor can always be written as a linear combination of two
linearly independent Majorana spinors, $\psi_0$ and $\psi_1$, as
\[\psi = \psi_0 + i \psi_1 \qquad \Rightarrow \qquad \Psi = \Psi_0 + {\cal I} \Psi_1.\]
Since $\bP {\cal I}={\cal I} \bP$,  $\bT {\cal I} = -{\cal I}\bT$ and $\bC {\cal I} = -{\cal I}\bC$, if
$\bC$ commutes with both $\bP$ and $\bT$ on Majorana fermions then it
also commutes when acting on Dirac spinors.
Hence demanding that $\bC$ commutes with $\bP$ and $\bT$ on Majorana spinors is sufficient to deduce that they commute when acting on Dirac spinors too, and $n=m=2$ giving $Q\times {\mathbb Z}_2$.

The possible parity eigenstates of Dirac fermions can also be deduced. Since $\bP \bPsi_1 = \pm {\cal I} \bPsi_1$ and $\bP \bPsi_2 = \pm {\cal I} \bPsi_2$ of the four possibilities
\begin{align*}
  \bP (\bPsi_1 + {\cal I}  \bPsi_2) & =  {\cal I}(\bPsi_1 + {\cal I}  \bPsi_2) ,\\
  \bP (\bPsi_1 + {\cal I}  \bPsi_2) & =  {\cal I}(\bPsi_1 - {\cal I}  \bPsi_2) ,\\
  \bP (\bPsi_1 + {\cal I}  \bPsi_2) & =  {\cal I}(-\bPsi_1 + {\cal I}  \bPsi_2) ,\\
  \bP (\bPsi_1 + {\cal I}  \bPsi_2) & =  -{\cal I}(\bPsi_1 + {\cal I}  \bPsi_2) ,  \end{align*}
only the first and fourth are eigenstates, hence
\[ \bP \bPsi = \pm {\cal I} \bPsi\]
and the parity is again $\pm i$. In the particle date tables protons and neutrons
are assigned parity $+1$, but this is a modified parity which is discussed in the conclusions.

Although $\bP^\dagger\ne - \bP$ with $\etaP=\pm i$, $\bP^\dagger \numeq -\bP$ is numerically anti-hermitian. Indeed
\[  \bP^\dagger \numeq - \bP, \qquad \bT^\dagger \numeq - \bT, \qquad  \Inv^\dagger \numeq - \Inv,
\qquad \bC^\dagger \numeq -\bC.\] 
Despite the fact that  $\bC {\cal I} = - {\cal I} \bC$ (and $\bC^\dagger \numeq - \bC$)
$\bC$ is actually unitary with $\bm \beta_-$.
To see this observe that
\[\bC^\dagger {\bm \beta}_\pm =\mp {\bm \beta}_\pm \bC,\] 
leading to
\[ \widehat \bC \bigl( \overline \bPsi \bPsi'\bigr)
:=\overline{\bigl(\bC \bPsi \bigr)} \bC \bPsi' 
= \overline{\bPsi}  \bPsi'.\] 
We also have
\begin{align} 
\\
\widehat \bC \bigl( \overline \bPsi \Gamma_5 \bPsi'\bigr)
&:=\overline {\bigl(\bC \bPsi\bigr)} \bC \Gamma_5 \bPsi' 
=- \overline {\bPsi} \Gamma_5  \bPsi', 
\\ 
\widehat \bC \bigl( \overline  \bPsi  \Gamma^\mu \bPsi'\bigr)
&:=\overline {\bigl(\bC \bPsi\bigr)}  {\bm \Gamma}^\mu \bigl(\bC \bPsi' \bigr) 
=-\overline {\bPsi}\bm \Gamma^\mu  \bPsi',
\end{align}
and, with $\bm \beta_+$,
\begin{align}
 \widehat \bC \bigl( \widetilde  \bPsi  \bPsi'\bigr)&:=\widetilde{\bigl(\bC \bPsi\bigr)} \bigl(\bC \bPsi' \bigr) 
=-\widetilde {\bPsi}  \bPsi'\\
 \widehat \bC \bigl( \widetilde  \bPsi  \Gamma_5 \bPsi'\bigr)&:=\widetilde{\bigl(\bC \bPsi \bigr)}  {\bm \Gamma}_5 \bigl(\bC \bPsi' \bigr) 
=\widetilde {\bPsi}\bm \Gamma_5  \bPsi'\\
\widehat \bC \bigl( \widetilde  \bPsi  \Gamma^\mu \bPsi'\bigr)
&:=\widetilde {\bigl(\bC \bPsi\bigr)}  {\bm \Gamma}^\mu \bigl(\bC \bPsi' \bigr) 
=\widetilde {\bPsi}\bm \Gamma^\mu  \bPsi'. \end{align}

\section{The CPT theorem}

The $\CPT$ theorem for Dirac fermions follows as usual.  We have
\[ \bTheta =  \CPT = \bpm e^{i(\phi_T+\phi_C-\phi_P)} \gamma_5 & 0 \\ 
0 & e^{-i(\phi_T+\phi_C-\phi_P)}(\gamma_5)^* \epm.\]
With
$ \phi_T + \phi_C=\frac{2\pi}{p}$ and
$p=1,2$, $\phi_P=\pm \frac \pi 2$ for $\varepsilon =+1$ while $p=4$,
$\phi_P=0,\pi$ for $\varepsilon =-1$,
this gives
\[ \bTheta =  \CPT = \pm i\,{\bm \Gamma}_5.
\]
and $\bTheta^2=-1$, $\bTheta^\dagger \numeq -\bTheta$.
We also have
\[ \bTheta^\dagger \bm \beta_\pm = \bm \beta_\pm \bTheta,\]
and the choice of $\varepsilon$ does influence these properties of $\Theta$.

$\bTheta$ changes the sign of both bi-linear forms, 
\beq
\overline {\bigl(\bTheta \bPsi\bigr)} \bigl(\bTheta \bPsi'\bigr) = 
-\overline {\bPsi}  \bPsi', \quad 
\widetilde {\bigl(\bTheta \bPsi\bigr)} \bigl(\bTheta \bPsi'\bigr) = 
-\widetilde {\bPsi}  \bPsi'.\label{eq:Thetaminus}\eeq
However $\widehat \bTheta$ interchanges the order of the fermions,
because $\widehat \bT$ does, and so changes the sign
\beq
\widehat \bTheta \bigl(\overline { \bPsi} \bPsi'\bigr) = 
\overline {\bPsi'}  \bPsi, \quad 
\widehat \bTheta \bigl( \widetilde  \bPsi \bPsi' \bigr) = 
\widetilde {\bPsi'}  \bPsi.\label{eq:Thetahatminus}\eeq
We have the following bi-linear relations (omitting the argument $x$)
\begin{align}
\widehat  \bTheta \bigl( \overline \bPsi \bPsi'\bigr) 
&= \overline {\bPsi'}  \bPsi \\
\widehat \bTheta \bigl( \overline \bPsi\, \bm {\bm \Gamma}_5 \bPsi'\bigr) 
&= \overline {\bPsi'}\bm {\bm \Gamma}_5  \bPsi  \label{eq:Theta5}\\
\widehat \bTheta \bigl( \overline \bPsi \,\bm \Gamma^\mu \bPsi'\bigr) 
&= -\overline {\bPsi'}\,\bm \Gamma^\mu  \bPsi\\
\widehat \bTheta \bigl( \overline  \bPsi\, {\bm \Gamma}_5 \bm \Gamma^\mu \bPsi'\bigr) 
&= -\overline  {\bPsi'}\,{\bm \Gamma}_5 \bm \Gamma^\mu \bPsi. 
\end{align}
Of course
$\widehat \bTheta$ also sends $x$ to $-x$ in $\bPsi(x)$ and
$\partial_\mu\rightarrow - \partial_\mu$, 
so
\beq
\widehat \bTheta \bigl(\overline   \bPsi {\cal I}\Gamma^\mu \partial_\mu\bPsi)
= \overline {\bPsi} {\cal I} \bm \Gamma^\mu  \partial_\mu \bPsi.\eeq
and the Dirac Lagrangian is invariant under $\widehat\bTheta$.

There is another bi-linear invariant that can be constructed
on $\spinor$, namely
\[\overline \psi  \psi'_c = e^{i\varphi_c}\, \psi^T  C^{-1}  \psi'\]
where $\psi'_c=\widehat C \psi'$ is the spinor conjugate to $\psi'$.
For a Majorana spinor $\psi'_c=\psi'$ and this is equal to $\overline \psi \psi'$.
The corresponding bi-linear on $\Spinor$ is
\[ \overline \bPsi  \bC  \bPsi'
  = e^{i\varphi_C} \bigl( \psi^\dagger  (C^*)^{-1}\psi'^*\bigr) - e^{-i\varphi_C} (\psi^T C^{-1} \psi ')\]
since, from  (\ref{eqa:bCbC}), $\beta C \beta^* = (C^*)^{-1}$.
For a Majorana spinor $\bPsi = \bC \bPsi$ equation
 (\ref{eq:PsibarPsi}) shows that 
 \[  \overline \psi \psi' + \overline {\psi'} \psi =e^{i\varphi_C} \bigl( \psi^\dagger  (C^*)^{-1}\psi'^*\bigr) - e^{-i\varphi_C} (\psi^T C^{-1} \psi ')\]
 and for both sides to be real we must have $e^{i\varphi_C}=\pm i$.
 Since
 \[e^{i(\varphi_T+\varphi_C)}=
  \pm \sqrt{\varepsilon}\]
 we conclude that
 $e^{i\varphi_T}=\pm \sqrt{-\varepsilon}$ also, so this argument
 determines all three phases factors in $\bP$, $\bT$ and $\bC$,
 up to a sign,
 $e^{i\varphi_P}=\pm \sqrt{-\varepsilon}$,  $e^{i\varphi_T}=\pm \sqrt{-\varepsilon}$ and  $e^{i\varphi_C}=\pm i$
 for Majorana spinors.

\section{Neutral pseudo-scalar mesons\label{sec:B-mesons}}

Experimentally the most fruitful place to study the properties of $\widehat \bC$, $\widehat \bP$ and $\widehat \bT$ is in the physics of pseudo-scalar mesons, particularly $B$-mesons for $\widehat \bT$.  
Pseudo-scalars are represented by operators of the form $\overline{\sPsi}\Gamma_5 \sPsi\,'$.
For example if $\sPsi_b(x)$  creates a $\bar b$-quark
and $\sPsi_d(x)$ creates a $\bar d$-quark\footnote{In the Fock space operator
  in equation (\ref{eq:Psi-Fock}), $a_s$ annihilates a fermion and $a^\dagger_{c,s}$ creates the anti-fermion.} then
\begin{align*} B^0&= \overline{\sPsi}_d \gamma_5 \sPsi_b |0>\\
\overline {B^0}&= \overline{\sPsi}_b \gamma_5 \sPsi_d |0> = \bigl(\overline{\sPsi_d} \gamma_5 \sPsi_b \bigr)^\dagger|0>
 \end{align*}
are the neutral $B$-mesons which are $\widehat P=-1$ eigenstates.
For a relativistic bound state, like the $B^0$, $\psi_b$ here need not be a single particle operator, the $\bar b$-quark will be accompanied by a sea of particles and antiparticles, but for the purposes of this section all that matters is that $\psi_b$ has the same Lorentz transformation properties of
a singe particle (\ref{eq:D-omega-psi}), even if it is a composite operator.

Now consider  $\overline{\bPsi}_d \bm \Gamma_5 \bPsi_b$ with $\bm \beta_-$ re-scaled by a factor of $\/\sqrt{2}$ for convenience,
\begin{align*} \overline{\bPsi}_d \bm \Gamma_5 \bPsi_b&=
     \frac {1}{\sqrt 2}
\bigl(\sPsi_d^\dagger, \sPsi_d^T\bigr) \bpm \beta & 0 \\ 0 & -\beta^* \epm \bpm \gamma^5 & 0 \\
0 & \gamma_5^* \epm \bpm \sPsi_b \\ \sPsi_b^\Dagger \epm\\
 & = \frac {1}{\sqrt 2}
 \bigl(\sPsi_d^\dagger \beta \gamma_5 \sPsi_b - \sPsi_d^T \beta^* \gamma_5^* \sPsi_b^\Dagger\bigr)
 =  \frac {1}{\sqrt 2}\bigl(\overline \sPsi_d \gamma_5 \sPsi_b + \sPsi_b^\dagger \gamma_5^\dagger \beta^\dagger \sPsi_d\bigr)\\
 & = \frac {1}{\sqrt 2}\bigl(\overline \sPsi_d \gamma_5 \sPsi_b - \overline{\sPsi}_b \gamma_5  \sPsi_d\bigr),
\end{align*}
since fermions anti-commute and $\gamma_5^\dagger \beta^\dagger = - \beta \gamma_5$.
Acting on the vacuum 
\[ \overline{\bPsi}_d \bm \Gamma_5 \bPsi_b |0> = \frac {1}{\sqrt 2}(B^0 -\overline{B^0})=B_+.\]
This is an eigenstate of both $\widehat \bP$, with $P=-1$ and $\widehat \bC$, with $C=-1$ so 
$C P=+1$.

The $C P=-1$ state is constructed using $\frac{1}{\sqrt{2}}\bm \beta_+$,
\[  \widetilde {\bPsi}_d  \bm \Gamma_5 \bPsi_b
=  \frac {1}{\sqrt 2}\bigl(\overline \sPsi_d \gamma_5 \sPsi_b + \overline{\sPsi}_b \gamma_5  \sPsi_d\bigr)\]
and
 \[\widetilde{\bPsi}_d  \bm \Gamma_5 \bPsi_b |0> = 
\frac {1}{\sqrt 2}\bigl( B^0 + \overline{B^0} \bigr)=B_-\]
is an eigenstate of both $\widehat \bP$, with $P=-1$ and $\widehat \bC$, with $C=+1$, so
this is the other $\widehat \bC\widehat \bP$ eigenstate with  $C P=-1$.

In the same way eigenstates of $\widehat  \bT$ can be constructed\footnote{For an eigenvectors 
of the matrix $\bT$ the component in $\spinor^*$ is not the complex conjugate of the component in $\spinor$.} and,
while $\bP$ and $\bT$ do not commute when acting on fermions, $\widehat\bT$ and $\widehat \bP$ commute when acting
on fermion bi-linears so neutral $B$-mesons can be simultaneous eigenstates of 
$\widehat \bC$, $\widehat \bP$ and $\widehat \bT$.
From (\ref{eq:Tminus}) and (\ref{eq:Tminus5})
\begin{align*}
  \widehat \bT\bigl(  {\overline \bPsi}_d  {\bPsi}_b\bigr) 
&= \overline {\bPsi}_b \bPsi_d \\
  \widehat \bT \bigl( \overline {\bPsi}_d  \bm \Gamma_5  {\bPsi}_b\bigr) 
&= \overline {\bPsi}_b \bm \Gamma_5\bPsi_d, 
\end{align*}
while $\bm \beta_+$ gives the opposite signs
\begin{align*}
  \widehat \bT \bigl(\widetilde {\bPsi}_d  {\bPsi}_b\bigr) 
&= -\widetilde {\bPsi}_b \bPsi_d \\
  \widehat \bT \bigl(\widetilde {\bPsi}_d \bm \Gamma_5 \bPsi_b\bigr) 
&= -\widetilde {\bPsi}_b \bm \Gamma_5\bPsi_d. 
\end{align*}
Thus $\overline {\bPsi}_d \bm \Gamma_5\bPsi_b$ has $C P=1$, $T=1$ while
$ \widetilde {\bPsi}_d \bm \Gamma_5\bPsi_b$ has $CP=-1$, $T=-1$.

In terms of $\bTheta=\CPT$
\begin{align} 
 \widehat \bTheta \bigl(\overline {\bPsi}_d  \bm \Gamma_5   \bPsi_b\bigr) 
&=  \overline {\bPsi}_b \bm \Gamma_5\bPsi_d\label{eq:ThetaGamma5}\\
 \widehat \bTheta \bigl(\widetilde{\bPsi}_d \bm \Gamma_5 \bPsi_b\bigr) 
&=  \widetilde{\bPsi}_b \bm \Gamma_5\bPsi_d
\end{align}
from which it is immediate that $CPT=1$ for both $B_+$ and $B_-$, as expected from the
 $\bC\bP\bT$ theorem.

$\bT$-violation has been observed in neutral $B$-meson systems by the BaBar collaboration \cite{TV-BaBar}. It would be interesting if the $\bT$-parity could be measured experimentally and compared to the $\CPT$ prediction.

\section{Discussion\label{sec:discussion}}

It has been shown that, assuming $\bP^{2 n} =1$ when acting on Dirac fermions for some integral $n$,
$\bP$ and and $\bT$ generate one of  an infinite series of possible groups, the generalised quaternion  groups.  
Including charge conjugation and assuming that $\bC$ commutes with
$\bP$ and $\bT$  singles out the unique option
\beq
\Inv^2 =\bP^2=\bT^2 = \Inv \bP \bT =-1, \qquad [\bC,\bP]=[\bC,\bT]=0. 
\label{eq:Broombridge}\eeq

Although the discussion in \S\ref{sec:PT} and \S\ref{sec:CPT} involved only single particle states in $\Spinor$
it is easily extended to multiparticle states involving products of 
$\Spinor$ in the usual way.

If $\bP$ and $\bT$ preserve the Majorana property of a Majorana spinor then the inclusion of $\bC$ eliminates all but one of these possibilities leaving two copies of the quaternion algebra as the only 
option. While this is perhaps a mathematically pleasing observation it is difficult to think of any experimental consequences.  

Before 1956 it was generally believed that $\bP$ was a symmetry of the fundamental laws of Nature,
and indeed it is for electromagnetism and the strong nuclear force (the r\^ole of $\bP$ in any putative quantum theory of gravity is still open to debate).
When Lee and Yang realised in 1956 that $\bP$ invariance had not been tested in weak interactions \cite{Lee+Yang} they discussed possible experiments to check its status
and $\bP$-violation was observed in $\beta$-decay of Co$^{60}$
very soon afterwards \cite{PV}, leading to the important conclusion that the Hamiltonian giving rise to $\beta$ decay does not commute with the parity operator. 
It was almost immediately suggested
by Landau that, if $\bP$ was replaced by $\bC \bP$, then $\bC\bP$ still seemed to be a good symmetry.  If so
 this would imply that $\bC$ is also violated (but in a manner that exactly cancels $\bP$-violation). 
Then $\bC \bP$ violation was discovered in the decay of neutral Kaons to pions 
in 1964 \cite{CPV},
\[ K_L \ \rightarrow \ 2\pi\]
where $K_L$ is $\bC\bP$ odd and the $2\pi$ state is $\bC\bP$ even 
(the latter is $s$-wave due to angular momentum conservation, since Kaons and pions are both spin-$0$).  So $\bC$-violation does not exactly cancel $\bP$-violation.  Famously $\bC\bP$ violation is essential to explain the
fact that there is more matter than anti-matter in the Universe \cite{Sakharov},
though the Standard Model alone does not have large enough $\bC\bP$ violation
to explain the observed matter/anti-matter asymmetry.
Nevertheless the $\bC \bP \bT$ theorem, discovered a few years prior to these events 
\cite{Luders1} - \cite{Bell}, suggests that the combination of all three operations
$\bC$, $\bP$ and $\bT$ should be a good symmetry of Nature.

The $\CPT$ theorem, together with $\bC \bP$ violation, implies that $\bT$ should also be violated,  but this is a theoretical prediction based on a particular model of fundamental interactions (assuming a local, Lorentz invariant Lagrangian)
 and should be tested against observations.
Experimental evidence for $\bT$-violation in neutral Kaon systems was claimed 
in \cite{TV-CPLEAR} and later in \cite{TV-Fermilab}. 
It was confirmed more definitively in a beautiful experiment 
by the BaBar collaboration \cite{TV-BaBar}, following a proposal in \cite{BM-VV-P}. In the chain of $\Upsilon(4S)$-decays
\begin{align*}
 \hbox{I:} \hskip 50pt  \Upsilon &\longrightarrow 
\begin{cases} 
 B^0 & \rightarrow  \quad  l^+ +X \\  
 {\bar B}^0 & \rightsquigarrow  \quad
 \begin{cases}
   B_- \quad  \rightarrow\quad  J/\Psi + K_S\\
   B_+ \quad  \rightarrow\quad  J/\Psi + K_L \end{cases}
\end{cases}  
\\
 \hbox{II:} \hskip 50pt \Upsilon &\longrightarrow 
\begin{cases} 
  B^0 & \rightsquigarrow  \quad  \begin{cases}
    B_-  \quad \rightarrow  \quad   J/\Psi + K_S\\
    B_+  \quad \rightarrow  \quad   J/\Psi + K_L\end{cases}\\
  \bar B^0 & \rightarrow \quad l^- + \overline X ,\\  
\end{cases} 
\end{align*}
the wavy arrows represent mixing rather than decays. $B^0$ and $\overline B^0$ 
are flavour eigenstates while $B_\pm$ are $\bC\bP$ 
eigenstates
and $l^\pm$ are charged leptonic states.\footnote{It
is essential for the experiment that,  in decay sequence I, $B^0$ and $\bar B^0$ form an entangled state: $\bar B^0$ is inferred 
from $B^0\rightarrow l^+ + X$ and it is not observed directly.  If it were observed it could not oscillate to the $C P$ eigenstate $B_-$. 
Similarly, in sequence II, $B_+$ and $B_-$ are an entangled state and $B_-$ is not observed directly.}  
There are 8 possible decay sequences falling into 4 pairs of time reversed oscillations, such as
\begin{align*}
l^+\,X \ \mbox{followed by }\  J/\Psi\,K_L &\\
  J/\Psi\, K_S  \ \mbox{followed by }\  l^-\,\overline X&
\end{align*}
for example: in this sequence the semi-leptonic decay $l^+ X$ can only come from a $B^0$ so
the subsequent $CP$ even state,
$J/\Psi\, K_L$,
must be due to a $B_+$ decaying, having oscillated from a $\bar B^0$,
($\bar B^0 \ \rightsquigarrow  \ B_+$).
The $l^- \overline X$ can only come from a $\bar B^0$ which has oscillated from a $B_+$ ($B_+ \ \rightsquigarrow \bar B^0$) because the earlier
$CP$ odd $J/\Psi\, K_S$ state must have come from a $B_-$. 
Experimentally the time of the two oscillations is different and,
  since the decays involved are $\bC\bP$ conserving, this asymmetry can only be due to  $\bT$-violation.

$\CPT$ invariance, which has a strong theoretical foundation,  replaces the now discredited ideas
of $\bP$ invariance and $\bC\bP$ invariance. It predicts that it is four-dimensional inversion $x \rightarrow -x$, combined with the interchange of particles with anti-particles, that is a symmetry of Nature.
This is very natural: the definitions of $\bP$ and $\bT$ require choosing a reference frame and, in a relativistic setting, they should be combined 
into $\Inv=\bT\bP$.  A relativistic statement of the $\CPT$ theorem would be the  $\bC\Inv$ theorem: that $\bC$-violation is exactly cancelled by 
$\Inv$ violation.  Landau would have got it right if only he had thought 
relativistically, as Bell did in \cite{Bell} where $\Inv$ was emphasised over $\bP$ and $\bT$ separately.  

Assuming that $\bP$ and $\bT$ commute with $\bC$ leads to the conclusion that
that a fermionic eigenstate of parity must have $\bP^2=-1$, so its parity is $\pm i$.
Particle data tables however give the parity of protons and neutrons as $+1$.
This is because when there is
a global $U(1)$ symmetry, such as baryon number ${\bm B}$, a new parity operator
\[\bP' = e^{i \vartheta_B {\bm B}}\bP \]
can be defined and $\vartheta_B$ chosen to ensure that $\bP'^2=1$, \cite{Weinberg}.
If $\bP$ is conserved so is $\bP'$, since the $U(1)$ is a symmetry, and
by convention $P'=+1$ is chosen for baryons. If $\bP$ is replaced with $\bP'$ the quaternion group is replaced by the dihedral group $D_8$, realising the second
group identified in \cite{Socolovsky}.

Some insight into the possible phase of $\bP$ (rather then $\bP'$) can be obtained by a short calculation along the lines of \cite{B+D}, page 41, 
 which shows that reflecting a plane-wave solution of the Dirac equation
$\psi_{inc}(t,z)$, with wavevector $\bm k =  k \hat {\bm z}$,  off
an infinitely high potential barrier ($V_0 \rightarrow \infty$ in \cite{B+D}), followed by a rotation of $\pi$
about the $z$-axis, produces a reflected wave 
$\psi_{ref}=i\gamma^0_{Dirac}\psi_{inc}(t,-z)$, suggesting that indeed $e^{i\phi_P}=i$ and $n=2$.

One might try to measure $\phi_P$ in a two slit experiment in which one path is subjected to reflection in a mirror followed by a rotation of $\pi$ about an axis perpendicular to the mirror
while the other path is left alone (the rotation could be induced by applying an appropriate magnetic field).  This is equivalent to $\bP$ acting on one path but not the other and a non-zero phase $\phi_P$ might then be expected to show up as a shift in the interference pattern. However this is probably not possible as reflecting the fermion off a mirror will inevitably impart a momentum to the mirror, no matter how small, which will make it possible, at least in principle, to determine which slit the fermion has gone through and presumably this will destroy the
interference pattern, even if the recoil of the mirror is not measured.
Certainly, if $\bP^2=-1$, then $\bP$ is not hermitian\footnote{Explicitly
  \[ \bigl(\overline {\bP \bPsi} \bigr) \bPsi = \overline \bPsi \bP^{-1} \bPsi \] and $\bP$ is only hermitian if $\bP^{-1} = \bP$. It is obviously anti-hermitian if
  $\bP^2 = -1$.} and it cannot be an observable. $\bP$ acting on fermions is not a physical observable, though acting on fermion bi-linears it is.
Indeed the whole quaternion algebra (\ref{eq:Q-alebra}),
while mathematically precise and elegant,
does not seem to be physically observable. When acting on fermion bi-linears
it reduces to ${\mathbb Z}_2\times{\mathbb Z}_2\times{\mathbb Z}_2$.

There are however physical consequences of $[\bC,\bP]=[\bC,\bT]=0$.
Weinberg \cite{Weinberg} argues that  $e^{i\phi_P}=\pm i$ for Majorana fermions.
Starting from (\ref{eq:Psi-Fock}), with $a_s=a_{c,s}$, he concludes that
$e^{i\phi_P}=-e^{i\phi_{c,P}}$. where $\phi_P$ is the phase for $\sPsi$
and $\phi_{c,P}$ is the phase for the conjugate spinor $\sPsi_c$.
For a Majorana spinor $\phi_{c,P} = \phi_P$ so necessarily $\phi_P = \pm \frac{\pi}{2}$.
But the explicit form (\ref{eq:Psi-Fock}) is derived from the assumption that
the 2-point correlator $\Delta(x-y) = \{\sPsi(x),\overline{\sPsi}(y)\}$
vanishes for space-like separations -- this assumption fixes the relative phases of the positive and negative energy terms in (\ref{eq:Psi-Fock}), or equivalently, the relative phases of $a_s$ and $a_{c,s}$. We can turn this argument around and say that assuming $\bC$ commutes with $\bP$ and $\bT$ implies
that the 2-point correlator of anti-commuting fields vanishes for space-like separations, implying an intimate correlation between complex conjugation of wave-functions and causal structure.  This is perhaps not too surprising when one considers that the chiral representations $D_\pm(\omega)$ of the Lorentz group
acting on Dirac spinors are faithful representations of $Sl(2,{\mathbb C})$ with a complex parameter ${\bm \theta} + i {\bm \alpha}$ (see (\ref{aeq:D-omega})), clearly the appearance of $i$ here is related to the causal structure.
If $\bC$ did not commute with the improper Lorentz group the causal structure
could be violated.

One final remark on time-reversal. In the $\Upsilon$ decays described above $\bT$-violation was deduced as a consequence of two processes,
 $\bar B^0 \rightsquigarrow B_+$ and $B_+ \rightsquigarrow \bar B^0$, happening in opposite orders. 
For $\bP$ and $\bT$ the order also matters, not only do $\bP$ and $\bT$ not commute with the Hamiltonian, but in general they do not commute with each other.
However while $\bP$ and $\bT$ do not commute when acting on fermions they do commute when acting on fermion bi-linears so it probably not possible to see this non-commutativity experimentally. However we can hope to measure $\bT$ on fermion
bi-linears. Neutral pseudo-scalar mesons, 
such as Kaons and neutral $B$-mesons can be eigenstates of
$\bP\bC$ with $CP=\pm 1$.
Since $\widehat T$ maps $\spinor$ to $\spinor^*$ it is not possible for a fermion in $\spinor$ to be an eigenstate of $\widehat T$, but a state in $\Spinor$ can be an eigenstate of $\bT$.
Since all three of $\bC$, $\bP$ and $\bT$ are mutually commuting on fermion bi-linears, it should be possible to define the $\bT$ parity of such states, or equivalently their $\bTheta$ parity.
From the $\bC \bP \bT$ theorem their $\bT$-parity should be equal to
their $\bC\bP$-parity.

It is a pleasure to thank Denjoe O'Connor and Aiyalam Balachandran for helpful discussions.

\appendix
 
\section{Conventions \label{sec:conventions}}

To be specific we give conventions with metric signature
\beq \eta^{\mu \nu}=\begin{pmatrix} 1 & 0 & 0 & 0 \\ 0 & -1 & 0 & 0 \\ 0 & 0 & -1 & 0\\ 0 & 0 & 0 & -1\end{pmatrix}\label{eqa:signature}\eeq
and
\beq
\{ \gamma^\mu,\gamma^\nu\} =2 \eta^{\mu \nu}
\eeq
($\mu,\nu=0,1,2,3$). The opposite signature can be accommodated by sending
$\gamma^\mu \rightarrow i \gamma^\mu$, keeping $\gamma_5$ and $\beta$ fixed.

In the chiral representation
\begin{align}
\gamma^0 = 
\begin{pmatrix}
 0 &  \sigma^0 \\
 \bar \sigma^0  & 0
\end{pmatrix},
 \qquad
\gamma^i = 
\begin{pmatrix} 
0 & \sigma^i \\ 
\bar \sigma^i& 0
\end{pmatrix}
\end{align}
with $i=1,2,3$ and $\sigma^i$ the Pauli matrices (numerically $\bar \sigma^i = - \sigma^i$). 
The chirality matrix is
\beq 
\gamma_5 = -i  \gamma^0\gamma^1\gamma^2\gamma^3=
\begin{pmatrix}
 {\bm 1} & 0 \\ 
0 & - \bar{\bm 1} 
\end{pmatrix}.
\eeq
It is standard to  write $\sigma^0=\bar \sigma^0$and $\bar \sigma^i = -\sigma^i$, but this can
be confusing as they act on different spaces.
When convenient we shall write
\[\sigma^0 \numeq \bar \sigma^0, \qquad \bar \sigma^i \numeq  -\sigma^i\]
meaning the  matrices are numerically equal but have different spinor index structure.
Thus the identity matrices acting on the different Weyl sectors, denoted ${\bm 1}$ and $\bar {\bm 1}$, satisfy
${\bm 1} \numeq \bar {\bm 1}$ but ${\bm 1} \ne \bar {\bm 1}$.
In the text, when it is deemed necessary for clarity,  $\lq\lq="$ will be referred to as spinorial equality and $\lq\lq\numeq"$ as numerical equality
(for definiteness numerical equalities are given for signature
(\ref{eqa:signature})).
In particular
\[ (\sigma^\mu)^\dagger = \sigma^\mu, \qquad 
(\bar\sigma^\mu)^\dagger = \bar\sigma^\mu, \qquad \bar {\bm 1} = {\bm 1}^\dagger.\] 
This may seem like an overly pedantic distinction but it proves to be an extremely
useful notation to keep track of the Lorentz transformation properties of spinors.

The generators of the Lorentz group in the spinor representation are
\[ J^{\mu \nu} = \frac i 4 [\gamma^\mu,\gamma^\nu]\]
and a general Lorentz transformation on spinors is 
\[ D(\omega) = e^{-\frac i 2 \omega_{\mu\nu}J^{\mu\nu}}\]
where $\omega_{0 i}=\alpha n_i$ is a boost in the ${\bm n}$ direction
with rapidity $\alpha$ and $\omega_{i j}=\epsilon_{i j k}\theta^k$ a rotation in the 
$i$-$j$ plane.
In the chiral basis $J^{\mu\nu}$ are block diagonal
\begin{align*}
J^{0 i} &= \frac i 2 \bpm -\tau^i & 0 \\ 
0 & \bar \tau^i \epm
:= \bpm J_+^{0 i} & 0 \\ 0 & J_-^{0 i} \epm \\
J^{i j } &=  \frac 1 2 \epsilon^{i j }{}_k  
\bpm \tau^k & 0   \\ 0 & \bar \tau^k \epm := \bpm J_+^{i j} & 0 \\ 0 & J_-^{i j} \epm
\end{align*}
with 
\begin{align*}
\tau^i &= -\frac 1 2 (\sigma^0 \bar \sigma^i- \sigma^i \bar \sigma^0)
= \frac i 2 \epsilon^i{}_ {j k } \sigma^j \bar \sigma^k\\
\bar \tau^i &= \frac 1 2 (\bar \sigma^0  \sigma^i- \bar \sigma^i  \sigma^0)
=\frac i 2 \epsilon^i{}_{j k }\bar \sigma^j \sigma^k,
\end{align*}
and
\[D(\omega)= 
\bpm 
e^{-\frac{i} {2} (\bm \theta - i\bm \alpha).\bm \tau} & 0 \\
 0 &  e^{-\frac{i} {2} (\bm \theta + i \bm \alpha).\bar{\bm \tau} }  
\epm.\]

As is well known the 4-dimensional Dirac representation
$J^{\mu \nu}$ is a reducible representation of the Lorentz group and decomposes into two 
inequivalent Weyl representations $J_+^{\mu\nu}$ and  $J_-^{\mu\nu}$, with
$ J_\pm^i =\frac{1}{2}\bigl(\epsilon^i{}_{j k}  J^{j k} \pm i  J^{0 i}\bigr)$
corresponding to $\tau^i$ and $\bar \tau^i$ respectively, and
\beq D_+(\omega) = e^{-\frac{i} {2} (\bm \theta - i\bm \alpha).\bm \tau} \qquad \hbox{and} \qquad 
D_-(\omega) = e^{-\frac{i} {2} (\bm \theta + i \bm \alpha).\bar{\bm \tau}}.\label{aeq:D-omega}\eeq
Denoting the space of Dirac spinors by $\spinor$ then
 $\gamma^\mu: \spinor \rightarrow \spinor$, $D(\omega): \spinor \rightarrow \spinor$ and $\spinor$ decomposes into the two eigenspaces of $\gamma_5$, $\spinor_\pm$,
with
\[\gamma_5:\spinor_\pm =\pm \spinor_\pm, \qquad 
D_+(\omega):\spinor_+ \rightarrow \spinor_+\qquad  \hbox{and} \qquad D_-(\omega):\spinor_-
\rightarrow \spinor_-.\]
Complex conjugation interchanges these two inequivalent representations, up to an equivalence $D_\pm^*(\omega) \sim D_\mp(\omega)$, 
\begin{align*}
D_+^*(\omega)& = e^{\frac{i} {2} (\bm \theta + i \bm \alpha).{\bm \tau}^*} =  e^{-\frac{i} {2} (\bm \theta + i\bm \alpha).{(-\bm \tau}^*)} = e^{-\frac{i} {2} (\bm \theta + i\bm \alpha).(\bar \epsilon \bar{\bm \tau}\bar \epsilon^{-1}) }= \bar\epsilon D_-(\omega) \bar \epsilon\,^{-1}\\
D_-^*(\omega)& = e^{\frac{i} {2} (\bm \theta - i\bm \alpha).\bar {\bm \tau}^*} =  e^{-\frac{i} {2} (\bm \theta - i\bm \alpha).{(-\bar{\bm \tau}}^*)}
 =e^{-\frac{i} {2} (\bm \theta - i\bm \alpha).(\epsilon {\bm \tau} \epsilon^{-1}) }= \epsilon D_+(\omega) \epsilon^{-1},
\end{align*}
with $\epsilon = i \tau^2$ and $\bar \epsilon = -i \bar \tau^2$. Thus 
the reducible representations $D(\omega)$ and $\{D(\omega)\}^*$ are also equivalent, 
\beq \{D(\omega)\}^* = {\mathscr C}^{-1} D(\omega) {\mathscr C} \qquad \mbox{with} \qquad {\mathscr C}
=\bpm 0 & \epsilon^{-1} \\
\bar \epsilon\,^{-1} & 0 \epm. \label{eq:script-C}
\eeq

If $\Psip \in \spinor_+$ and $\Psim \in \spinor_-$, then a Dirac spinor $\sPsi$
decomposes into two Weyl spinors as
\beq \sPsi =\bpm \Psip \\ \Psim\epm. \eeq
In the dreaded dotted and undotted spinor notation, with spinor indices $a=1,2$ and $\dot a=1,2$, this is
\beq \bpm (\psi_+)_a \\ (\psi_-)^{ \dot a} \epm, \label{aeq:psi-chi-xi}\eeq
and the components of $\sigma^0$, $\bar \sigma^0$, $\sigma^i$, $\bar \sigma^i$, $\tau^i$ and 
$\bar \tau^i$ are
\[ \delta_{a \db},\quad
\delta^{\da b},\quad
(\sigma^i)_{a \db}, 
 \quad -(\sigma^i)^{\da b},
 \quad (\tau^i)_a{}^b,
\hbox{and} \quad (\bar \tau^i)^{\da}{}_\db,
\]
respectively. Numerically 
\[\sigma^i \numeq -\bar\sigma^i \numeq \tau^i \numeq \bar\tau^i \]
while $\bar \tau^i = (\tau^i)^\dagger$ is a spinorial equality. 
When multiplying matrices together in a manner that preserves 
Lorentz transformation properties a lower dotted index can only be contracted with an upper dotted index and a lower undotted index can only be contracted with an upper undotted index. Thus
$\sigma^\mu \bar \sigma^\nu$
is allowed but $\sigma^\mu \sigma^\nu$ is not,
though $\sigma^\mu \bar \sigma^\nu \numeq \sigma^\mu \sigma^\nu$.
The identity matrices ${\bm 1}$ and $\bar {\bm 1}$, acting on $\spinor_+$ and $\spinor_-$
respectively, have components
\[ \delta_a{}^b \qquad \hbox{and} \qquad \delta^{\da}{}_\db.\]
The product ${\bm 1} \bar {\bm 1}$ is not allowed and strictly speaking
the transpose matrix ${\bm 1}^{\tr}\ne {\bm 1}$.  Also $\sigma^0\numeq\bar \sigma^0 \numeq {\bf 1} \numeq \bar {\bm 1}$, but they should not be 
identified.  

Denote the vector space dual to $\spinor$ by $\spinor^D$, 
then an inner product on the space of Dirac spinors is defined between two 
spinors $\sPsi\in\spinor$ and $\sPsi\,'\in\spinor$ as
\[ \overline \sPsi \sPsi\,' = \sPsi^\dagger \beta \sPsi\,'\]
where $\sPsi^\dagger\beta \in\spinor^D$.  Then $\beta:\spinor \rightarrow \bigl(\spinor^D\bigr)^\dagger$ is an hermitian Lorentz invariant metric, $\beta=\beta^\dagger$,
satisfying 
\[ D^\dagger(\omega) \beta D(\omega) = \beta\]
which acts on $\gamma$-matrices as
\beq \beta\gamma^\mu = \bigl(\gamma^\mu\bigr)^\dagger \beta.\label{aeq:beta-gamma--dagger}\eeq
Explicitly 
\[ \beta= \bpm  0 & \bar {\bf 1} \\ {\bf 1} & 0 \epm \qquad \hbox{and} \qquad 
\beta^{-1} = \bpm  0 & {\bf 1} \\ \bar {\bf 1} & 0 \epm,\]
so
\beq \overline \sPsi \sPsi' = 
\psi_+^\dagger \psi_-' + \psi_-^\dagger \psi_+' =
(\psi_+^*)_\da (\psi_-')^{\da} + (\psi_-^*)^a  (\psi_+')_a.
\label{aeq:psibar-psi'}\eeq
and, 
when $\sPsi'=\sPsi$,
\beq \overline \sPsi\sPsi
=  
\psi_+^\dagger \Psim + \psi_-^\dagger \Psip.
\label{aeq:psibar-psi}\eeq

The matrices $\beta$, $\beta^{-1}$, and $\beta^*$ are not spinorially equal, though numerically
\[\beta\numeq \beta^{-1}\numeq \beta^*,\qquad \beta^2\numeq 1
\qquad \hbox{and} \qquad\beta \numeq \gamma^0 \numeq  \bigl(\gamma^0\bigr)^\dagger.
\]
However
\beq\beta = \beta^\dagger\eeq 
is correct.

There are also spaces $\spinor^D_+$ dual to $\spinor_+$ 
and $\spinor^D_-$ dual to $\spinor_-$ with
$(\psi_-^\da)^\Dagger = (\psi_-^*)^a $, $(\psi_{+\, a})^\Dagger = (\psi_+^\Dagger)_\da$
and $\psi_-^*\in \spinor^D_+$ and $\psi_+^\Dagger\in \spinor^D_-$.  So
\[\Dagger:\spinor_- \rightarrow \spinor^D_+, \qquad \Dagger:\spinor_+ \rightarrow \spinor^D_-.\]
Lorentz invariant metrics can be defined on $\spinor_+$ and $\spinor_-$ separately.  Let 
\[ \bigl(\psi_+^D\bigr)^a = \epsilon^{a b}(\psi_+)_b \qquad 
\hbox{and} \qquad  \bigl(\psi^D_-\bigr)_\da = \bar \epsilon_{\da \db} \psi_-^\db\]
with \[\epsilon\numeq\bpm 0 & 1 \\ -1 & 0 \epm 
\qquad (\mbox{components} \ \epsilon^{a b})\]
and 
\[\bar \epsilon \numeq\bpm 0 & -1 \\ 1 & 0 \epm\qquad  (\mbox{components}\  \bar\epsilon_{\da \db}).\]
Then $\epsilon$ is a metric on $\spinor_+$ and $\bar\epsilon$ a metric on $\spinor_-$.
These are Lorentz invariant metrics in that
\[D^\tr_+(\omega)\, \epsilon \, D_+(\omega)=\epsilon, \qquad 
D^\tr_-(\omega)\, \bar \epsilon \,D_-(\omega)=\bar\epsilon.\]
The inner product of $\Psip$ with itself is
\[ (\Psip)_a  \epsilon^{a b} (\Psip)_b = \psi_+ \psi_+^D=\psi_+^D \psi_+,\] which would vanish if $(\Psip)_a$ were commuting objects, but if they anti-commute (either as second quantised 
single particle fields or as Grassmann variables in first quantisation) then the above definitions are consistent.

The inverse metrics are
\[\epsilon^{-1}\numeq\bpm 0 & -1 \\ 1 & 0 \epm \qquad (\mbox{components} \ \epsilon_{a b})\]
and 
\[\bar \epsilon\,^{-1} \numeq \bpm 0 & 1 \\ -1 & 0 \epm
\qquad (\mbox{components} \ \bar \epsilon\,^{\da \db}).\]
This follows general relativistic conventions where it is standard to write the
components of the inverse of the metric $\bar \epsilon_{\da \db}$ as $\bar\epsilon\,^{\da \db}$,
so $\bar\epsilon_{\da \dc} \bar\epsilon\,^{\dc \db}  = \delta_\da{}^\db$, and 
$\epsilon^{a c} \epsilon_{c b}  = \delta^a{}_b$.
Just as in general relativity, spinor indices are raised and lowered by using these metrics (multiplying from the left, if multiplying from the right their 
transposes must be used).
In index free notation
\begin{align} 
\epsilon &= -\epsilon^T\label{eq:epsilon-relation1}\\
\bar \epsilon &= -\bar \epsilon^\tr\label{eq:epsilon-relation2}\\
\epsilon^{-1}\epsilon &= {\bf 1}\label{eq:epsilon-relation3}\\
\epsilon\epsilon^{-1} &= {\bf 1}^\tr\label{eq:epsilon-relation4}\\
\bar \epsilon\bar \epsilon\,^{-1} &= \bar {\bf 1}\label{eq:epsilon-relation6}\\
\bar \epsilon\,^{-1}\bar \epsilon &= \bar {\bf 1}^\tr\label{eq:epsilon-relation5}\\
\epsilon^*&=\bar \epsilon\,^{-1}\label{eq:epsilon-relation7}\\
\bar \epsilon^* &= \epsilon^{-1}.\label{eq:epsilon-relation8}\end{align}
While numerically 
\begin{align*}
\epsilon \numeq \epsilon^*  \numeq -\epsilon^{-1}& \numeq -\bar \epsilon
\numeq  -\bar \epsilon^* \numeq \bar \epsilon\,^{-1} 
\\
 \epsilon^2 & \numeq \bar \epsilon^2
\numeq -{\bm 1}.\end{align*}

The parity, time reversal and charge conjugation operators,  
$\HatP$, $\HatT$ and $\HatC$ respectively, are defined in the usual way 
in second quantisation (see {\it e.g.} \cite{Weinberg} with some minor changes due to different conventions)
\begin{align}
\HatP\sPsi(x) =\, &\etaP \gamma^0 \sPsi(x_P)\numeq \etaP \beta \sPsi(x_P)
\label{aeq:P}\\
\HatT\sPsi(x) =\, & \etaT C^* \beta \gamma_5 \gamma^0 \sPsi(x_T)\numeq -\etaT C \gamma_5 \sPsi(x_T)\label{aeq:T}\\
\HatC\sPsi(x) =\, &\etaC C \beta^\tr \sPsi^\Dagger(x)\numeq \etaC C \beta \sPsi^\Dagger(x)\label{aeq:C}
\end{align}
where the charge conjugation matrix transposes the $\gamma$-matrices,
\beq 
\gamma^\mu C = -C\bigl(\gamma^\mu\bigr)^\tr, \label{aeq:Cgamma-transpose}
\eeq
$\etaP$, $\etaT$, $\etaC$ are arbitrary complex phases, $x_P=(t,-{\bm x})$ and 
$x_T=(-t,{\bm x})$.
Explicitly
\beq C= \bpm \epsilon^{-1}  & 0 \\ 0 & \bar \epsilon\,^{-1} \epm 
\numeq \bpm -\epsilon  & 0 \\ 0 & \epsilon  \epm  \eeq
and
\[ {\mathscr C}= C \beta^T\]
in (\ref{eq:script-C}), so
\[ \widehat C \{ D(\omega) \psi \}
= e^{i\phi_C}{\mathscr C} \{D(\omega)\}^* \psi^*
= e^{i\phi_C}  D(\omega){\mathscr C} \psi^*.\]

From (\ref{eq:epsilon-relation7}) and (\ref{eq:epsilon-relation8}),
\[ C^* = \bpm \bigl( \epsilon^{-1} \bigr)^* & 0 \\ 0 & 
\bigl(\bar  \epsilon\,^{-1} \bigr) ^* \epm
= \bpm \bar \epsilon & 0 \\ 0 & \epsilon \epm, \]
with
\[ C^\tr  = -C, \quad  C^\dagger = -C^*,\quad C^* \numeq C\numeq -C^{-1},
\quad C^2\numeq -1 \quad  \hbox{and} \quad  C^\dagger C \numeq {\bm 1}.\]
It is natural that $\gamma^0$ should appear in the definitions of $\HatP$ and $\HatT$
as these require choosing a time-like direction and planar space-like foliations perpendicular to the chosen time direction: there is no such information in $\beta$, which is Lorentz invariant.

Clearly $\HatP : \spinor \rightarrow \spinor$ and $\widehat C:\spinor \rightarrow \spinor$ interchange $\spinor_+$ and $\spinor_-$,
while the charge conjugation matrix $C:\bigl(\spinor^D\bigr)^T \rightarrow \spinor$.
Also $\HatT:\spinor_\pm  \rightarrow \spinor^\Dagger_\pm $.  

In practice $\widehat C$, $\widehat P$ and $\widehat C \widehat T$ 
are easily implemented using the numerical equalities in 
(\ref{aeq:P})-(\ref{aeq:C}),
\begin{align}
\HatP\sPsi(x) &\numeq \etaP \bpm 0 & 1 \\ 1 & 0 \epm \sPsi({\cal P}x)
\label{aeq2:P}\\
\HatT\sPsi(x) &\numeq \etaT \bpm \epsilon & 0 \\ 0 & \epsilon \epm  \sPsi({\cal T}x)\label{aeq2:T}\\
\HatC\sPsi(x) & \numeq  \etaC \bpm 0 & -\epsilon \\ \epsilon & 0 \epm \sPsi^\Dagger(x).\label{aeq:C2}
\end{align}
Since
$\HatP:\spinor_\pm \rightarrow \spinor_\mp$,
$\HatT :\spinor_\pm \rightarrow \spinor_\pm^* \sim  \spinor_\mp$ and
$\HatC :\spinor_\pm \rightarrow \spinor_\mp$,
none of these discrete symmetry operators can be represented
on an irreducible representation of the Lorentz group.

The anti-particle of an anti-particle is the particle,
\[ \widehat C (\widehat C \sPsi) = \sPsi,\]
so $\widehat C^2=1$.

If $\sPsi$ and $\sPsi'$ are two fermions in $\spinor$ then
\[ \overline{(\HatC \sPsi)} (\HatC \sPsi') = \sPsi^\tr \beta^* C^\dagger\beta C \beta^* \sPsi'^\Dagger=-\sPsi^\tr \beta^* \sPsi'^\Dagger=\overline {\sPsi'} \sPsi,\]
since spinors anti-commute and $\beta^\dagger=\beta$ so $\beta^*=\beta^\tr$. 
This uses the important identity
\beq
C \beta^*  C^* \beta = 1 \label{aeq:Cbeta-relation} 
\eeq
which is easily proven by using (\ref{aeq:beta-gamma--dagger}) and 
(\ref{aeq:Cgamma-transpose}) to show that
\[ C \beta^* C^* \beta \gamma^\mu = \gamma^\mu C \beta^* C^* \beta \]
and the only matrix that commutes with all four $\gamma^\mu$ is a multiple of the identity matrix, so
\[ (C \beta^*) (C^* \beta)=\lambda {\bf 1}. \]
Sine the eigenvalues of $C \beta^*$ and $C^* \beta$ have complex conjugate eigenvalues $\lambda$ must be a positive real number and, since from (\ref{aeq:beta-gamma--dagger}) and (\ref{aeq:Cgamma-transpose}) the eigenvalues of
$\beta$ and $C$ have unit modulus, $\lambda=1$.
In particular
\[\overline{(\HatC \sPsi)} (\HatC \sPsi)= \overline{\sPsi} \sPsi,\]
with the physical interpretation that anti-particles have the same mass as particles. 

For $\HatT$ there is no explicit complex conjugation in (\ref{aeq:T}) yet $\HatT \sPsi(x) \in \spinor^\Dagger$, it is the matrix
$C^* \beta$ that maps $\spinor$ to $\spinor^\Dagger$.  The complex conjugate of (\ref{aeq:T}) is
\begin{align}\bigl\{\HatT\sPsi(x)\bigr\}^\Dagger 
& = \etaTstar C \beta^* \gamma_5^* \bigl(\gamma^0\bigr)^* 
\sPsi^\Dagger(x_T)\in \spinor \label{aeq:Tstar}\\
\Rightarrow \qquad \widehat T\bigl\{ \widehat T \sPsi(x) \bigr\}^\Dagger &
 = C^* \beta \gamma_5 \gamma_0 C \beta^* \gamma_5^*(\gamma^0)^* \sPsi^*(x)=-\sPsi^*(x).
\end{align} 
using (\ref{aeq:Cbeta-relation}) and (\ref{aeq:Cgamma-transpose}).
The metric on $\spinor^*$ is $\beta^*$ so we have the inner product
\begin{align} 
(\HatT \sPsi)^\dagger \beta^* (\HatT \sPsi')
&=
\sPsi^\dagger \bigl(\gamma^0\bigr)^\dagger \gamma_5^\dagger \beta^\dagger C^\tr 
\beta^*
C^* \beta \gamma_5  \gamma^0  \sPsi' \nonumber \\
& = -\sPsi^\dagger(\gamma^0)^\dagger \gamma_5^\dagger \beta \gamma_5 \gamma^0 \sPsi' \nonumber\\
&= \overline{\sPsi}\sPsi',\label{aeq:Tminussi-Unitary}
\end{align}
because $\gamma_5^\dagger \beta = -\beta \gamma_5$.

Since $\widehat T$ maps $\spinor \rightarrow \spinor^*$ we cannot apply $\widehat C$ in (\ref{aeq:C}) to $\widehat T \psi$ directly, but it can be 
applied to $\{\widehat T \psi\}^*$ in (\ref{aeq:Tstar}),
\begin{align*}
 \widehat C{\bigl\{\widehat T \psi\bigr\}^*} 
&= e^{i\phi_C} C \beta^T \{e^{-i\phi_T} C \beta^* \gamma_5^* (\gamma^0)^* \psi^* \}^*\\
& = e^{i(\phi_C+\phi_T)} C \beta^T C^* \beta \gamma_5 \gamma^0\psi\\
&=e^{i(\phi_C+\phi_T)}\gamma_5 \gamma^0\psi.
\end{align*}
This is in $\spinor$ and so we can apply $\widehat P$ directly
to obtain 
\[  \widehat P\widehat C {\bigl\{\widehat T \psi\bigr\}^*}=e^{i(\phi_P+\phi_C + \phi_T )}\gamma^0 \gamma_5 \gamma^0 \psi =-e^{i(\phi_P+\phi_C + \phi_T )} \gamma_5 \psi.  \] 

In the text any specific choice of $\gamma$-matrix representation is avoided as much as possible --- no physical result can depend on the choice of representation.
The definitions
\beq
\beta \gamma^\mu = \varepsilon (\gamma^\mu)^\dagger\beta,
\quad\gamma^\mu C = -C  (\gamma^\mu)^T, \label{eqa:betaC-def}
\eeq
implying
\beq C^*\beta \gamma^\mu = - \varepsilon (\gamma^\mu)^* C^* \beta,
\quad
C \beta^* C^* \beta = 1,  
\label{eqa:bCbC}\eeq
are sufficient to derive all the formulae in the text without choosing a representation for the $\gamma$-matrices.

\end{document}